\documentclass[%
 reprint,
 superscriptaddress,
 amsmath,amssymb,
 aps,
 pra,
 floatfix,
]{revtex4-1}

\usepackage{dcolumn}
\usepackage{bm}
\usepackage{gnuplottex}
\usepackage{tikz}
\usepackage{braket}
\usepackage{hyperref}
\usepackage{dsfont}
\usepackage{graphicx,enumerate,verbatim}
\usepackage{float,mathrsfs}
\usetikzlibrary{positioning,shapes}
\usepackage{xcolor}
\usetikzlibrary{arrows}
\usepackage{rotating}
\definecolor{babyblueeyes}{rgb}{0.63, 0.79, 0.95}
\definecolor{blue(pigment)}{rgb}{0.2, 0.2, 0.6}
\definecolor{darkerblue}{rgb}{0.0, 0.0, 0.4}
\definecolor{darkblue}{rgb}{0.0,0.0,0.5}
\definecolor{darkgreen}{rgb}{0.0,0.4,0.0}
\hypersetup{
   colorlinks,
   linkcolor=blue(pigment),
    citecolor=darkgreen,
   urlcolor=darkblue
}

\begin{document}
\newcommand{\note}[1]{\textcolor{blue}{(#1)}}
\newcommand{\Om}[1]{\small $\omega_{#1}$}
\newcommand{\De}[1]{$\Delta_{#1}$}
\newcommand{\Ga}[1]{$\Gamma_{#1}$}
\title{Rydberg Entangling Gates in Silicon}

\author{E. Crane}
 \email{e.crane@ucl.ac.uk}
 \affiliation{Department of Electrical Engineering and London Centre for Nanotechnology, University College London, Gower Street, London WC1E 6BT, United Kingdom}
\author{A. Schuckert}
\affiliation{Department of Physics, Technical University of Munich, 85748 Garching, Germany}
\author{N. H. Le}
\affiliation{Advanced Technology Institute and Department of Physics, University of Surrey, Guildford GU2 7XH, United Kingdom}
\author{A. J. Fisher}
\affiliation{Department of Physics and Astronomy and London Centre for Nanotechnology, University College London, Gower Street, London WC1E 6BT, United Kingdom}

\begin{abstract}
In this paper, we propose a new Rydberg entangling gate scheme which we demonstrate theoretically to have an order of magnitude improvement in fidelities and speed over existing protocols. We find that applying this gate to donors in silicon would help overcome the strenuous requirements on atomic precision donor placement and substantial gate tuning, which so far has hampered scaling. We calculate multivalley Rydberg interactions for several donor species using the Finite Element Method, and show that induced electric dipole and Van der Waals interactions, calculated here for the first time, are important even for low-lying excited states. We show that Rydberg gate operation is possible within the lifetime of donor excited states with 99.9\% fidelity for the creation of a Bell state in the presence of decoherence.
\end{abstract}

\maketitle

Donor electron spin qubits in silicon are competitive in most of Di Vincenzo's criteria for a quantum computer implementation: scalability (and CMOS compatibility), high-fidelity initialisation and readout as well as extremely long coherence times~\cite{Muhonen_2015,Tyryshkin2012, Morse_phosphorus_2018,LoNardo2015,Morse2017}.
%% of the order of seconds for electron or hyperfine spins of \textsuperscript{31}P or \textsuperscript{77}Se\textsuperscript{+}~\cite{Tyryshkin2012, Morse_phosphorus_2018,LoNardo2015,Morse2017}. 
%However, most scalable implementations of entangling gates rely on tuning a two-qubit phase $\phi = u \Delta t$, where $u$ is the two-qubit interaction strength, usually given by the exchange interaction, and $\Delta t$ is the time during which they interact~\cite{He2019,Borjans2020}. 
However, published experimental results of  entangling gates in silicon report only up to 86\%~\cite{He2019} or 90\%~\cite{Borjans2020} fidelity. Moreover, entangling exchange interactions can change by orders of magnitude if the distance between orbital ground-state spins is varied by one lattice site~\cite{Koiller2002}. Due to the impact on gate durations, this renders scaling challenging as even hydrogen lithography cannot to date perform donor placement accurate to within a single silicon lattice site~\cite{Fuechsle2012,Stock2020}.

\begin{figure}
\centering
\begin{picture}(250,180)
\put(15 ,70){\includegraphics[scale=0.18]{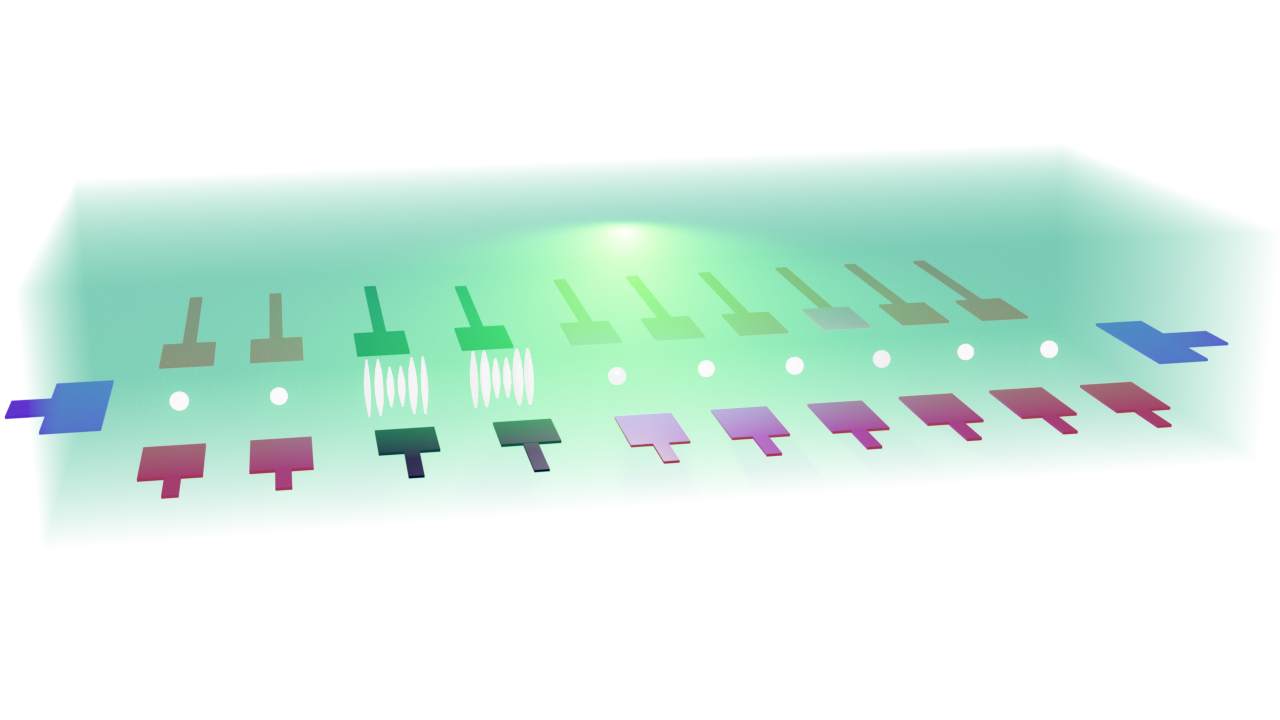}}
\put(11,0){\includegraphics[scale=0.12]{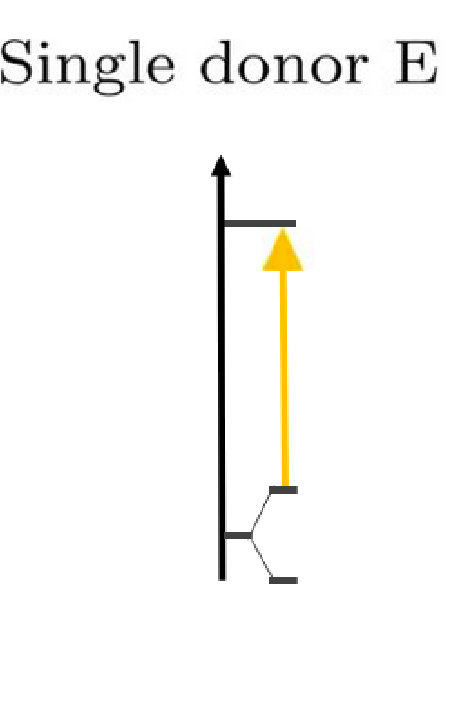}}
\put(80,-3){\includegraphics[scale=0.17]{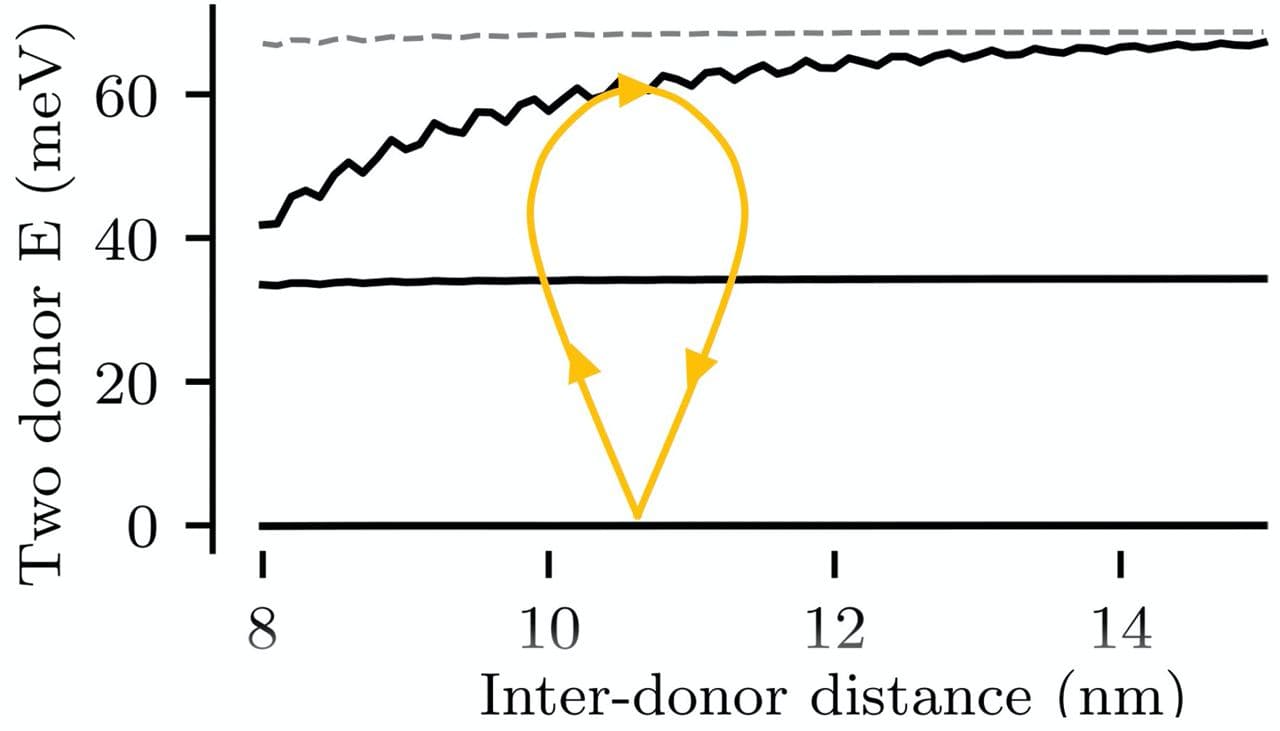}}
\put(0,170){a)}
\put(0,85){b)}
\put(67,85){c)}
\put(230,28){$\ket{gg}$}
\put(230,62){$\ket{rg}$}
\put(230,92){$\ket{rr}$}
\put(7,55){\textcolor{gray}{2p0} $\ket{r}$}
\put(7,18){\textcolor{gray}{1sA} $\ket{g}$}
\put(50,24){\textcolor{gray}{T0}}
\put(51,11){\textcolor{gray}{S0}}
\put(64,24){$\ket{1}$}
\put(64,11){$\ket{0}$}
\end{picture}
\caption{\label{fig_1}\textbf{Rydberg gate in silicon} a) Illustration of the Rydberg entangling gate silicon quantum computer  showing the global laser tuned to the $\ket{g}\rightarrow\ket{r}$ transition. The entangled qubit pair is selected by locally Stark shifting all other donors off resonance (red gates). Global gates (blue) enable use of induced dipolar interactions. Readout devices can be placed on the layer below the one shown here. This scheme can be extended for a surface code quantum computer similarly to \cite{Hill2015}. b) The ground state hyperfine levels are used as a qubit basis, with a higher lying orbital excited state as $\ket{r}$. c) The interactions between two atoms in the excited state barely oscillate, leading to a distance-robust two-qubit phase acquired after a Rabi cycle (residual oscillations are due to multivalley interference in the exchange interaction).}
\end{figure}
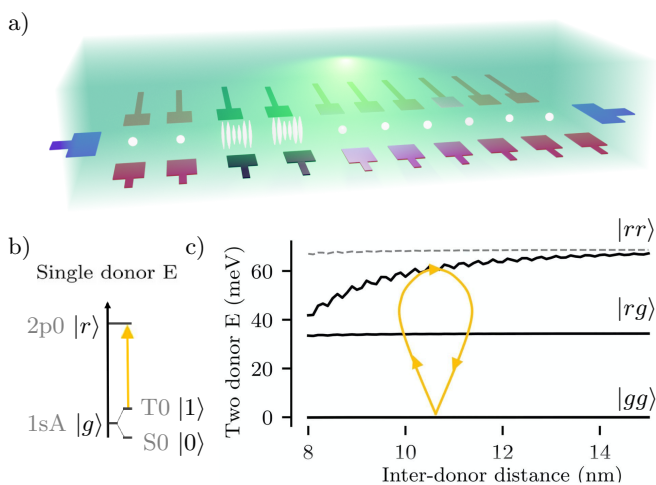

In this paper we draw on the analogy between donors trapped in the silicon lattice and alkali Rydberg atoms trapped in light fields~\cite{Greenland2010} to propose a new Rydberg entangling gate which is robust to variations in the inter-qubit distance, with theoretical fidelities up to 99.9\% in silicon. Our gate relies on the enhancement of the interactions $u$ between atoms in an orbital excited (\textit{Rydberg}) state $\ket{r}$ which leads to rapid gate operations on the order of hundreds of picoseconds. We show that the theoretical fidelities achievable by our new gate protocol are an order of magnitude higher than existing Rydberg atom schemes. This is because our gate allows for the excited state interaction $u$ to be on the same order of magnitude as the Rabi frequency $\Omega$ coupling one of the qubit levels $\ket{1}$ spin-selectively (see Fig.~\ref{fig_1}) to $\ket{r}$.

Specifically, we show here that while in cold atoms excitation to \emph{high-lying} excited states is needed to induce strong dipole interactions, for donors in silicon the interactions generated by the \emph{low-lying} excited states are large enough due to the large dielectric screening and small effective mass in silicon. We calculate for the first time electric (not magnetic~\cite{Sousa_Magnetic_2004}) dipole and Van der Waals interactions between excited states. The Finite Element Method (FEM) enables quick and accurate calculations of large numbers of silicon donor wavefunctions of high lying excited states (with or without an electric field) within the effective mass approximation, required for perturbation theory. To predict the fidelities we simulate the entangling gate operation in the presence of a finite excited state lifetime $T_1$, assuming the decoherence time to be twice the lifetime, and show that our proposal outperforms existing Rydberg schemes. By using the experimentally determined $T_1$ and our numerically determined $u$, we find that high fidelities independent of placement errors can be reached in silicon. Lastly, we show that this gate is robust to variations in donor energy levels (inhomogeneous broadening) and that only tuning of a single independent parameter is needed.

\begin{table}
\centering
\begin{tabular}{c||c|c|c|} 
Donor & Se+ & Se+ & P/As \\
$\ket{r}$ & 2p0 & 1sT2 & 2p0 \\ 
\hline
$d_{\ket{1} \rightarrow \ket{r}}$ (Debye) & 0.97$^a$ & 1.96 $^b$ & 31 $^c$ \\ 
$\hbar \omega _{\ket{1} \rightarrow \ket{r}}$ (meV) & 548 & 427 & 34 \\ 
%\hline 
%$\hbar\Omega$ (meV) & 245 & 194 & 8600$^3$ \\ 
theo. $T_1$ (ns) & -$^d$ & - & 1.1/1.8 $^e$ \\ 
exp. $T_1$ (ns) & -$^d$ & 7.7$^f$ & 0.235$^g$ \\ 
exp. $T_2$ (ns) & - & 7.7$^h$ & 0.160$^j$ \\ 
%\hline 
%$h\gamma$ (meV) & 1$^4$ & 0.13$^5$ & 4.2$^6$ \\ 
\hline 
\end{tabular} 
\caption{\label{fig_2}\textbf{Properties of orbital excited states used in this work.} a. calculated using FEM results. b. From \cite{DeAbreu2019}. c. From \cite{Clauws_oscillator_1988}. Two-photon excitation via the D0X state could also be envisaged ($d_{\ket{1} \rightarrow \ket{D0X}}$=0.04 Debye and $d_{\ket{D0X} \rightarrow \ket{r}}$=0.003 Debye~\cite{Gullans2015}). d. We assume a conservative $T_1$=1ns. e. From \cite{Tyuterev_Theoretical_2010}. f. From \cite{DeAbreu2019}. Photo-luminescence radiative $T_1$ = 900 ns, however modulation-frequency-dependent luminescence direct $T_1$ = 7.7 ns~\cite{DeAbreu2019} (may be limited by thermal phonons leaving room for improvements in $T_1$ \cite{Royle2018}). g. From \cite{Hubers2013}. Dominated by phonon-assisted relaxation h.  From \cite{DeAbreu2019}. Limited by $T_1$. j) From \cite{Litvinenko2015}. The sample was not isotopically pure, leaving room for improvements in $T_2$.}
\end{table}

\textbf{Rydberg entangling gates.--} Whereas previous gate schemes~\cite{Lukin2001,Jaksch2000,levine_parallel_2019}  required $\hbar/T_1 \ll \Omega < u$, our proposal only requires $\hbar/T_1 \ll \Omega,u$, allowing for much smaller gate durations and hence less probability of decay from $\ket{r}$, leading to higher gate fidelities. This is because previous schemes were based on the blockade, in which the doubly excited state $\ket{rr}$ is tuned out off resonance due to strong interactions induced by $\Omega \ll u$. Therefore, the gate duration $\tau\sim 1/\Omega$ is limited to $\tau\gg 1/u$. In our protocol, due to $\Omega$ being on the same order of magnitude as $u$, the gate duration can be $\tau\sim 1/u$.

The pulse sequence (see supplement for details ~\cite{supplement}) consists of acting twice with a pulse with Rabi frequency $|\Omega|$ detuned from the $\ket{r}$ transition on both donors simultaneously for a duration chosen such that the first pulse returns $\ket{01}$ and $\ket{10}$ to themselves. The phase  $\xi$ of $\Omega$ is then chosen such that $\ket{11}$ returns to itself after a second pulse of the same duration and Rabi frequency $|\Omega|$. During the two pulses, $\ket{11}$ picks up a two-qubit phase due to the interactions in $\ket{rr}$. Lastly, the detuning $\Delta$ is chosen such that the final phases of all two-qubit states implement a controlled phase gate. Global addressing enables parallel entangling gate operations as demonstrated in Ref.~\cite{levine_parallel_2019}. Conditions for a successful gate implementation are: spin-selective excitation to $\ket{r}$ (either leaving the $\ket{0}-\ket{r}$ transition off resonance or forbidden by a selection rule) and $u, \Omega > \hbar/T_1$. In the following, we show that these two conditions can be fulfilled in both shallow and deep donors.%We calculate inter-donor interactions $u$ with the FEM and use experimentally determined excited-state lifetimes $T_1$.

\textbf{Spin-selective excitation to $\ket{r}$.--} 
The singly ionised deep double donor selenium (\textsuperscript{77}Se\textsuperscript{+}) has a large hyperfine interaction in the ground state, giving rise to singlet S0 and triplet T0 states as the qubit basis with $T_1$ beyond 6 minutes and $T_2$ beyond 2 s~\cite{Morse2017}. For $\ket{r}$ both the $\pm$ valley states of 1sT$_2\Gamma_7$~\cite{Morse2017} and 2p0~\cite{DeAbreu2019} can be used, as they are dipole-allowed spin-selective transitions with easily accessible excitation energies.

For the shallow donors phosphorus (P), arsenic (As), antimony and bismuth, the qubit can be encoded in the hyperfine S0 and T0 states with a T2 time exceeding 7s~\cite{Morse_phosphorus_2018}. For $\ket{r}$ we take the 2p0 state, which can be accessed from the D0X level~\cite{Gullans2015,Morse_phosphorus_2018} or using micromagnets to emulate spin-orbit coupling, as the larger extent of the $2p0$ wavefunction would lead to a larger Zeeman splitting in $\ket{r}$ than in $\ket{g}$~\cite{Pioro-Ladriere2007}. Higher lying states would give rise to larger $u$, at the cost of a smaller $\Omega$ or the need for a two-photon excitation, as is the case for cold atoms.

For all proposals, we have checked that $\Omega>h/T_1$ is feasible with laser intensities below 1 W/$\mu m^2$, and we list their parameters in table ~\ref{fig_2}.

Having discussed that spin-selective excitation of orbital-excited states is indeed possible, we now calculate the interactions between donors in $\ket{r}$ to find $u$.

\textbf{Rydberg interaction $u$.--} Entangling gate schemes so far have focussed on harnessing the highly oscillatory and exponentially decaying ground state ferromagnetic exchange ($J$)~\cite{He2019,Optically_Crane_2019}. This interaction is negligible in high-lying excited states of Rydberg atoms, where $u$ is dominated by the electric dipole interaction~\cite{Lukin2001}
\begin{equation}
V_{dd} = \frac{\bm{p}_1\bm{p}_2 - 3(\bm{n}\cdot\bm{p}_1)(\bm{n}\cdot\bm{p}_2)}{R^3},\label{eq_vdd}
\end{equation}
where $\bm{p_i}$ is the dipole operator of atom $i$, $R$ is the inter-atomic distance, and $\bm{n}$ is the unit vector pointing along the inter-donor axis. In the absence of an electric field, neutral atoms and donors do not have a permanent dipole moment so $V_{dd}$ only contributes at second order in perturbation theory: Van der Waals ($V_{V\!dW}$) interactions falling off as $\sim 1/R^6$, see Supplemental Material~\cite{supplement} whose strength is determined by the polarisability which scales with the principal quantum number as $n^{11}$. Contrarily, $J$ is determined by the size of the wavefunctions which scale as $n^2$ and decay exponentially with inter-donor distance. Therefore, dipolar interactions dominate for large principal quantum number and distance. 

In silicon, the six conduction band minima lead to donor electron multivalley wavefunctions. Here, we calculate multivalley results for $J$, electron-electron repulsion ($W$) and for the first time also $V_{V\!dW}$ and the induced dipole interaction $V_{dd}$ in the presence of an electric field. We employ the FEM to solve the Schr\"odinger equation exactly within effective mass theory:
\begin{equation}
\big[ -\frac{\hbar^2}{2 m^*}\nabla ^2 + U_{imp}(r) + U_{cc}(r) + U_{Ef}(r) \big] \psi_i (r) = \epsilon_i \psi_i (r),\notag
\end{equation}
where $U_{cc}$ is a central-cell potential for the S states which have a non-negligible probability of being at the core, $U_{Ef}$ is the electric field, $\epsilon_i$ is the binding energy of the donor and $m^*$ is the effective mass of the electron in the silicon lattice, see Supplemental Material~\cite{supplement}. Then we couple the result into the six valleys. The FEM solves for any number of eigenstates, enabling a convergence check of perturbative calculations. For $V_{V\!dW}$ interactions, we typically include $\sim 40$ states. Similarly, we obtain $V_{dd}$ from the dipole moment which we get from the slope of the multi-valley Stark shift. This perturbatively calculated Stark shift does not contain contributions from the continuum, which we quantify by comparing to the single valley energies directly generated by the FEM while varying $U_{Ef}$ (see Supplemental Material~\cite{supplement}). As the outermost donor electron is very weakly bound to the core, we set the maximally applicable electric field (which wouldn't ionise $\ket{r}$ during its lifetime) by generalizing the method in ref.~\cite{Yamabe1977} to donors in silicon, see Supplemental Material~\cite{supplement}, consistent with experimental results.

\begin{figure}
\centering
\begin{picture}(250,124)
\put(-2,4){\includegraphics[scale=0.15]{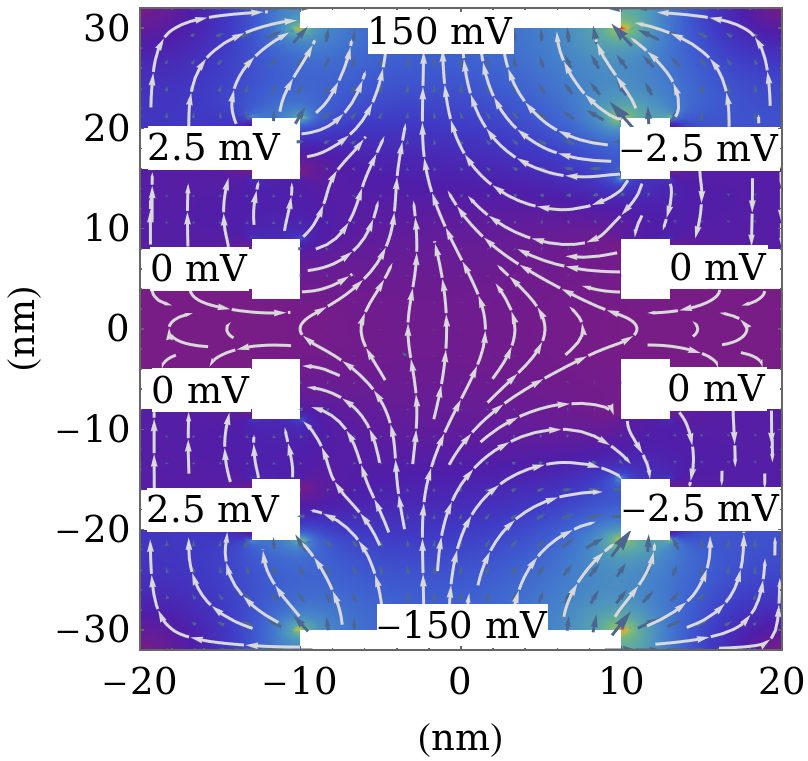}}
\put(-2,80){\begin{sideways}\tiny [0,0,1]\end{sideways}}
\put(80,5){\tiny [0,1,0]}
%\put(119,0){\includegraphics[scale=0.8]{P_2p0_lines_ar2.pdf}}
%\put(201,31){\includegraphics[scale=0.07]{l2}}
\put(123,30){\includegraphics[scale=0.12]{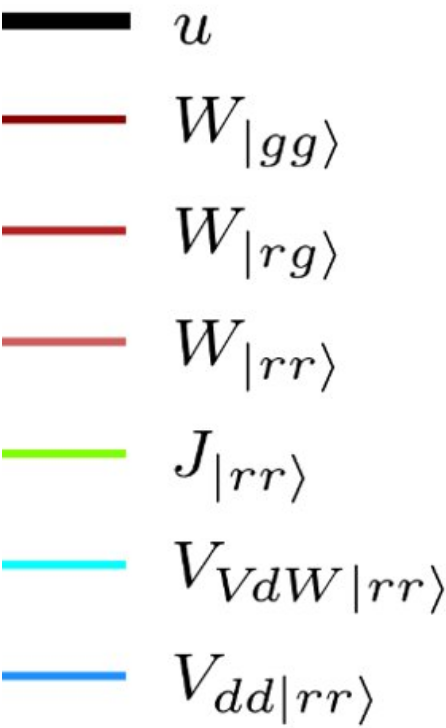}}
\put(165,0){\includegraphics[scale=0.8]{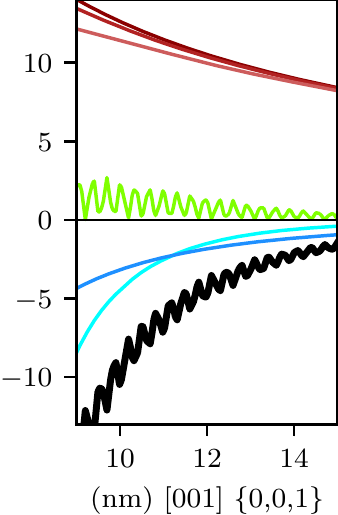}}
\put(0,112){\small a)}
\put(121,112){\small b)}
\put(162,59){\begin{sideways}\tiny (meV)\end{sideways}}
%\put(133,110){Si:P or Si:As 2p0-2p0}
\put(200,22){\includegraphics[scale=0.05]{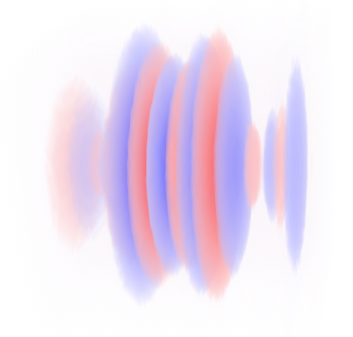}}
\put(224,22){\includegraphics[scale=0.05]{wavefunction2.png}}
\put(59,73){\includegraphics[scale=0.04,angle=90,origin=c]{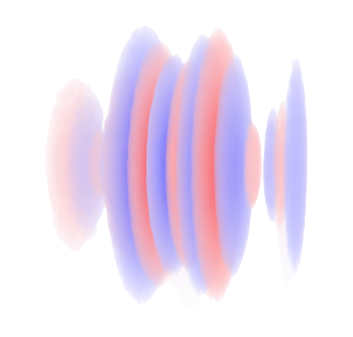}}
\put(59,52){\includegraphics[scale=0.04,angle=90,origin=c]{wavefunction2-removebg-preview}}
\put(62,90){\includegraphics[scale=0.025]{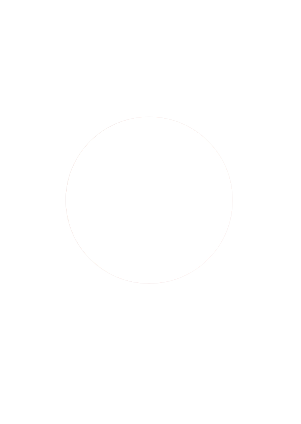}}
\put(62,35){\includegraphics[scale=0.025]{whitedot.png}}
\end{picture}
\caption{\label{fig_interactions}
\textbf{Electric fields and Rydberg  interaction energy ($u$).--} a) Electrostatic simulation of a device. Top and bottom global gates apply 0.18 V/$\mu$m (very low ionisation probability, see Supplemental Material~\cite{supplement}) between the two central donors, inducing dipolar interactions. The top and bottom donors are shifted off resonance with the laser due to sidegates applying a field of 0.5 V/$\mu$m (as in Fig.~\ref{fig_1}). Electric field from 0 to 3 V/$\mu$m. Polarisation parallel to inter-donor axis \{0,0,1\}. b) Interactions in meV between As or P donors in this same setup. $u$: total Rydberg interaction, $W$: electron-electron repulsion, $J$: ferromagnetic exchange, $V_{VdW}$: Van der Waals, $V_{dd}$: induced dipole interaction at 0.18 V/$\mu$m. `r' and `g' refer to Rydberg and ground state as defined in Fig.~\ref{fig_1}. 2p0 wavefunction illustration (blue: negative, red: positive), showing the absence of multi-valley oscillations perpendicular to the polarisation direction.}
\end{figure}

Fig.~\ref{fig_interactions} shows all interactions between two 2p0 state phosphorus donors and the relevant Rydberg interaction $u$ which corresponds to the additional energy experienced by the atoms when they are doubly excited as opposed to singly excited:
\begin{align}
u&=(E_{\ket{rr}}-E_{\ket{gr}})-(E_{\ket{rg}}-E_{\ket{gg}})\label{eq_ryd}\\
&\approx W_{\ket{rr}}-2W_{\ket{rg}}+W_{\ket{gg}}-J_{\ket{rr}}+V_{V\!dW\ket{rr}},\label{eq_rydexp}
\end{align}
where the subscript $\ket{g}$ denotes the ground state. $J_{\ket{gg}}, V_{V\!dW\ket{gg}}, V_{V\!dW\ket{rg}},J_{\ket{rg}}$ are negligible for the range of donor separations considered in this work.

%In the presence of a non-ionising electric field, $V_{dd}$ dominates $W$. 
Contrary to the expectation from the ground orbital state and despite the small principal quantum number, both $V_{V\!dW}$ and $V_{dd}$ interactions are non-negligible. When two donors are aligned along the polarisation and electric field axis as is shown in figure~\ref{fig_2}a), $V_{V\!dW}$ and $V_{dd}$ dominate, with an oscillatory contribution from $J$. When aligned along an axis perpendicular to the polarisation, $J$ and $V_{V\!dW}$ dominate, with no oscillations due to the $2p0$ state not showing multivalley oscillations in the plane perpendicular to the polarization axis (see Fig.~\ref{fig_interactions}c)). In the Supplemental Material~\cite{supplement}, we show that for the $1sT_2\Gamma_7$ state in Se+ the dipole interactions are less dominant and $u$ is mainly given by $J$.

For a successful gate operation in a quantum computer, the interactions $u$ need to be switchable witout affecting the surrounding qubits. Selecting the qubit pair to be excited to $\ket{r}$ requires Stark shifting the donors into resonance with the laser using electric fields~\cite{He2019,Watson2018}, analogously to proposals for Rydberg atoms in optical lattices~\cite{weiss_quantum_2017}. Single qubit operations can be done in the same way, using global microwave illumination~\cite{Watson2018}. We performed an electrostatic simulation of the gate configuration sketched in Fig.~\ref{fig_1}a), showing in Fig.~\ref{fig_interactions} that the electric gate configuration sufficient for shifting donors off-resonance would not create stray electric fields harmful to device operation. It would also be possible to apply an electric field roughly parallel to the inter-donor axis, enabling induced dipole interactions while avoiding ionising the donors, see Supplemental Material~\cite{supplement}.

In all donors and $\ket{r}$ studied here, $u$ is on the order of 1-10meV, fullfilling $u > \hbar/T_1$. Given a large enough laser power, the interaction scale sets the ultimate limit to gate operation times, enabling entangling gate durations of a fraction of a nanosecond, nine to ten orders of magnitude faster than the qubit lifetime.

In order to show that high-fidelity entangling gates are possible within the short donor lifetime, we calculate the fidelity of Bell state creation for various donors in the presence of decoherence by simulating the whole pulse sequence within a Lindblad Master equation approach in the following.

\textbf{Fidelity of Bell-state creation in presence of decoherence.--} We model a two donor system with the Hamiltonian~\cite{Jaksch2000} 
\begin{align}
H&=\sum_{i=1,2} \left(\frac{\Omega_i}{2} |1\rangle_i\langle r|+\frac{\Omega_i^*}{2} |r\rangle_i\langle 1|-\Delta|r\rangle_i\langle r|\right)\notag\\&+u\,|r\rangle_1\langle r|\otimes |r\rangle_2\langle r|,
\end{align} 
and simulate the full pulse sequence for creating the Bell-state $|\Phi^+\rangle=\frac{1}{\sqrt{2}}|00\rangle+|11\rangle$ within a Markovian Lindblad Master equation given by
\begin{equation}
    \partial_t \rho = -\frac{i}{\hbar} \left[H_{\mathrm{eff}}\rho-\rho H_{\mathrm{eff}}^\dagger\right]+\sum_j L_j \rho L_j^\dagger,
    \label{eq:Lindblad}
\end{equation}
where $\rho$ is the two-donor density matrix (with each donor restricted to the states $\ket{0}$, $\ket{1}$ and $\ket{r}$), $L_j$ are the jump operators describing decoherence and $H_{\mathrm{eff}}=H-\frac{i}{2}\sum_j L_j^\dagger L_j$. We include dephasing between $\ket{1}$ and $\ket{r}$ with rate $1/(2T_1)$ and spontaneous emission from the Rydberg state, see Supplemental Material~\cite{supplement}, with rate $1/T_1$. Decoherence processes between the qubit levels are negligible on the time scale of a single two-qubit gate.

We compare the Bell-state creation fidelity $\mathcal{F}=\bra{\Phi^+}\rho\ket{\Phi^+}$ between the previous schemes and our proposal, see Supplemental Material~\cite{supplement} and Fig.~\ref{fig_final}. According to the hierarchy of scales $\hbar/T_1\ll \Omega<u$ present in the previous schemes discussed above, there is a clear optimum for $\Omega$~\footnote{In the experiment in Ref.~\cite{levine_parallel_2019} $\Omega$ was not optimized, however it was close to  optimal one we find numerically, in part explaining their high fidelities.}. In the gate proposed in this paper, there is also an optimal $\Omega$ which we determine numerically along with the other pulse parameters (see supplement~\cite{supplement}). We find that the only parameter which requires tuning is $\Omega$, which depends linearly on $u$:
\begin{align}
\Omega/u&=1.45747,\\
\Delta/\Omega&=0.28757,\\
\xi&=1.5306.
\end{align}
For a given $u/\gamma$, the optimal Rabi frequency for our proposal is an order of magnitude larger than in the previous schemes, leading to much shorter gate durations and higher fidelities. We also find the fidelity to be robust to small variations in $\Omega$ and $u$. It also remains unchanged under realistic inhomogeneous broadening, which would lead to a variation of the excited state energies and therefore the detuning, see Supplemental Material~\cite{supplement}.

\begin{figure}
\centering
\begin{picture}(250,250)
\put(0,122){\includegraphics[scale=0.145]{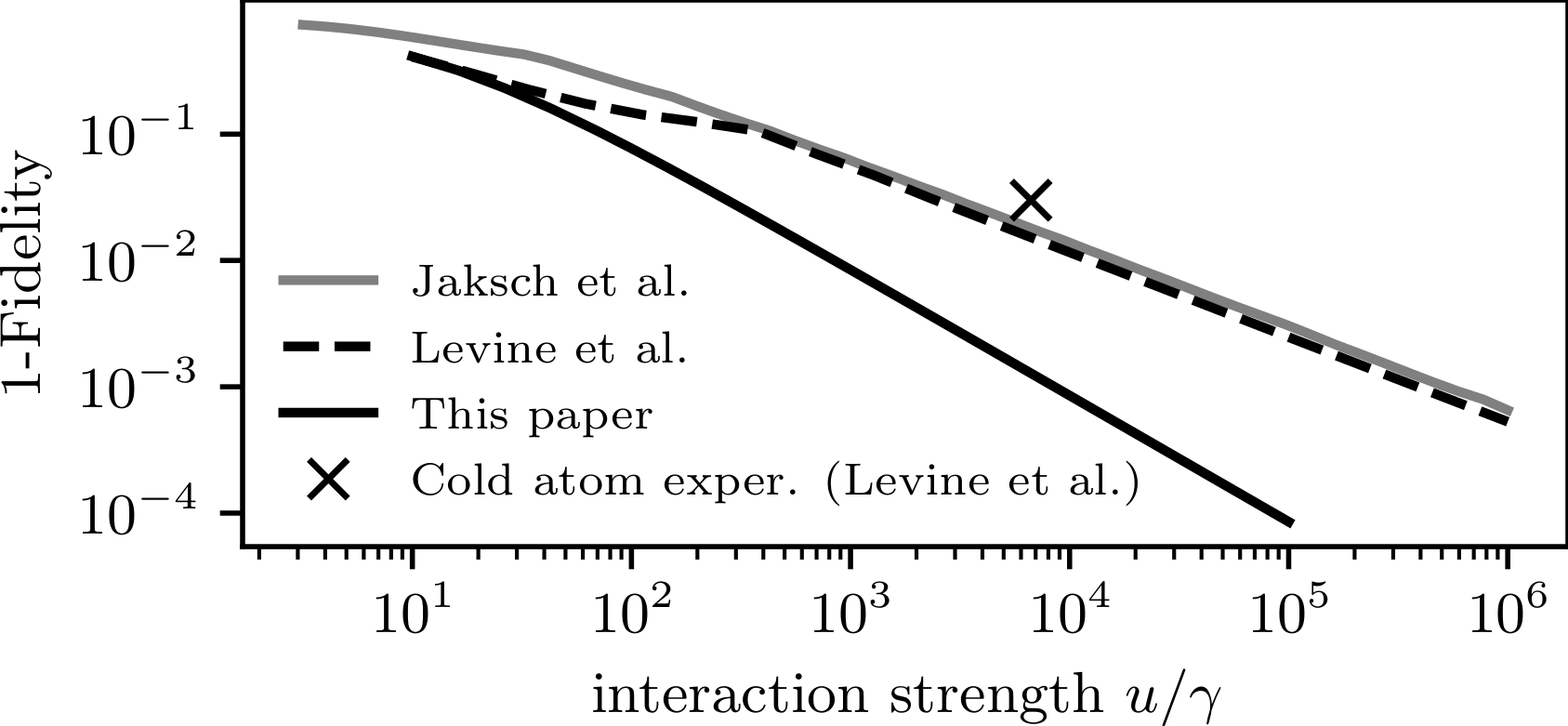}}
\put(24,0){\includegraphics[scale=0.136]{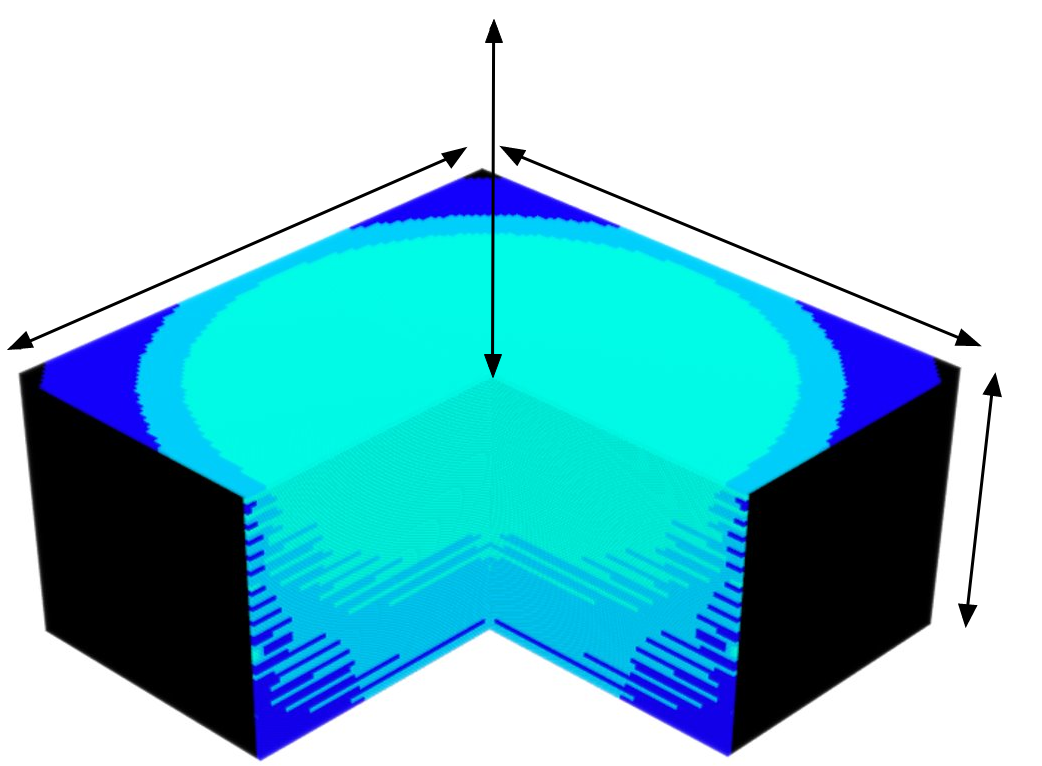}
%\includegraphics[scale=0.92]{block_fid-tdec.pdf}\quad
%\raisebox{10mm}[0pt][0pt]{\includegraphics[scale=0.32]{legf.png}}
}
\put(96,91){8 nm}
\put(37,67){\begin{rotate}{26}10.5 nm\end{rotate}}
\put(117,79){\begin{rotate}{-24}10.5 nm\end{rotate}}
\put(164,51){\begin{rotate}{-96}5.4 nm\end{rotate}}
\put(88,103){$\times$   donor qubit 1}
\put(19,96){polarisation}
\put(15,94){\vector(0,1){10}}
\put(19,79){Ef}
\put(15,79){\vector(0,1){10}}
\put(195,19){\includegraphics[scale=0.12]{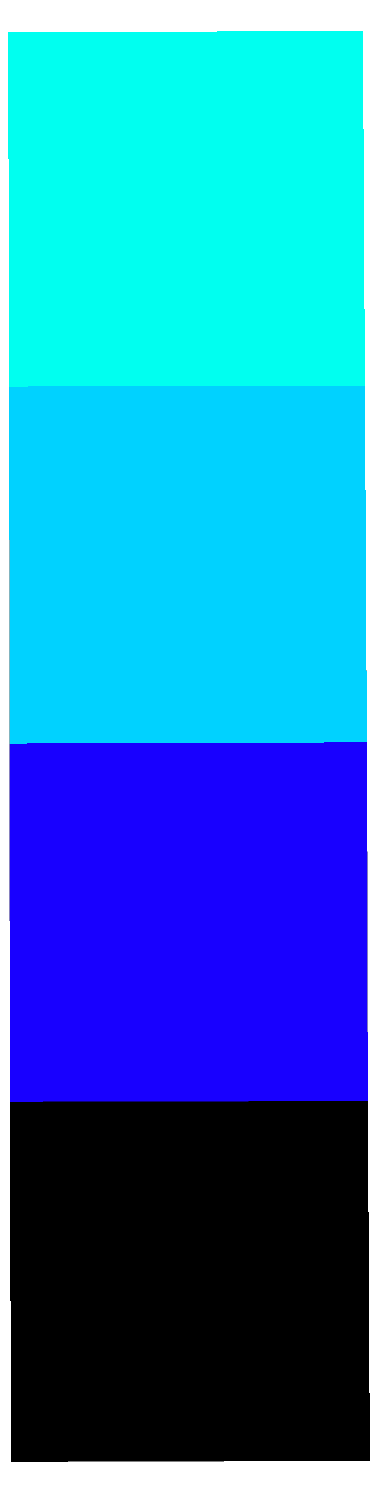}}
\put(195,76){Fidelity}
\put(210,63){$> 99.5 \%$}
\put(210,50){$> 99 \%$}
\put(210,37){$> 95 \%$}
%\put(0,0){\includegraphics[scale=1]{robustnessy.pdf}}
%\put(120,165){\vector(3,-1){30}}
%\put(120,165){\vector(-3,1){30}}
%\put(155,129){\vector(0,-1){19}}
%\put(155,129){\vector(0,1){19}}
\put(0,232){a)}
%%\put(160,175){\large \textbf{$\times$}}
%%\put(143,172){Si:Se+}
%%\put(135,189){\large \textbf{$\times$}}
%%\put(115,190){Si:P /}
%\put(110,182){Si:As}
\put(0,104){b)}
\end{picture}
\caption{\label{fig_final} \textbf{Predicted Bell-state creation fidelities for general Rydberg gates (a) and for Si:P or Si:As (b)} a) Results from the Lindblad simulation of the pulse protocols from Jaksch et al.~\cite{Jaksch2000} and Levine et al.~\cite{levine_parallel_2019} along with the protocol proposed in this paper and the experimental data on alkali atoms in optical tweezers~\cite{levine_parallel_2019,bernien_probing_2017}.  b) Fidelity map (plotted for 95\% and above) as a function of distance between donors shows that Fidelities are insensitive to donor placement. Experimental lifetime $T1=235$ps as shown in Tab.\ref{fig_2}. $\ket{r}$ is the 2p0 state polarised along \{0,0,1\}. The same fidelity region appears, diametrically opposed with regards to donor 1. 8nm minimum separation for avoiding the molecular limit in the 2p0 state~\cite{Li_Radii_2018}. Fidelity robust to donors belonging to different silicon sublattices (see Supplemental Material~\cite{supplement}) or diffusing out of plane (Si:As diffusion $<$1nm\cite{Stock2020}). For the corresponding Si:Se results, see Supplemental Material~\cite{supplement}.}
\end{figure}
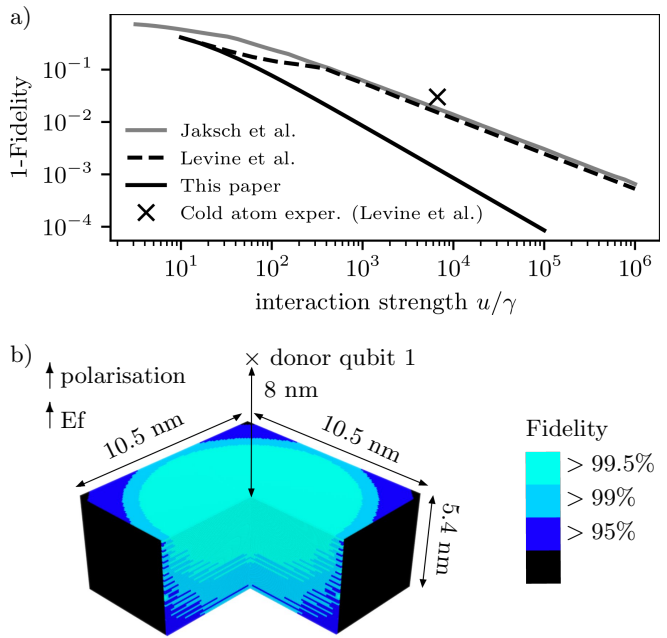

We use these results to predict the Bell-state creation fidelities possible in silicon by entering the interactions $u$ calculated in the previous section and the lifetimes summarized in Table ~\ref{fig_2}. In Fig.~\ref{fig_final}b) we show the gate fidelity as a function of the distance between donors. Importantly, fidelities of approximately 99.9\% can be reached. Moreover, the fidelity is only weakly dependent on distance such that placement errors as large as several nanometers make little difference and it does not make a difference if both donors reside on different sublattices of the silicon lattice (see supplement~\cite{supplement})

To conclude, we have proposed a new Rydberg gate, potentially increasing fidelities in cold atom experiments by an order of magnitude. Furthermore, we have shown that they can be implemented in donors in silicon, reaching fidelities of 99.9\% and having a high tolerance for donor placement error, opening the possibility to implement entangling gates in solid state quantum computers with ion implantation~\cite{ZIEGLER20101818}. The insensitivity to variations in the interaction strength and the inhomogeneous broadening not only brings advantages for scalability, but also reduces fine-tuning in device operation and may therefore be advantageous for other solid-state platforms. We emphasize that the physical nature of the interactions is irrelevant, in particular both exchange and dipolar interactions may be used. Acceptors do not have multivalley oscillations~\cite{Zhu_acceptor_2020} and would therefore be ideal candidates for this robust gate implementation as their dipolar interactions may be large enough to obtain high fidelities over extremely large areas, leading to a robust, long range entangling gate in the solid state.

\begin{acknowledgements}
We thankfully acknowledge discussions with R. Crane, S.
Ebadi, H. Pichler, N. J. Curson, B. N. Murdin, W. Wu and E. Orlova. We gratefully acknowledge financial support from the UK Engineering and Physical Sciences Research Council (COMPASSS/ADDRFSS, Grant No. EP/M009564/1). A.S. acknowledges financial support from the International Max Planck Research School for Quantum Science and Technology (IMPRS-QST) funded by the Max Planck Society (MPG).
\end{acknowledgements}

%\begin{picture}(250,250)
%\put(0,115){\includegraphics[scale=0.16]{Ef2.png}}
%\put(167,143){\includegraphics[scale=0.29]{label2}}
%\put(5,0){\includegraphics[scale=0.9]{P_2p0_lines_ar.pdf}}
%\put(0,240){a)}
%\put(0,100){b)}
%\put(133,110){Si:P or Si:As 2p0-2p0}
%%\put(175,200){\includegraphics[scale=0.11]{wavefunction2.png}}
%%\put(150,240){b)}
%%\put(165,240){2p0, pol \{0,0,1\}}
%%\put(210,230){$\rightarrow$}
%\put(82,25){\includegraphics[scale=0.06]{wavefunction2.png}}
%\put(106,25){\includegraphics[scale=0.06]{wavefunction2.png}}
%\put(167,25){\includegraphics[scale=0.06,angle=90,origin=c]{wavefunction2.png}}
%\put(202,25){\includegraphics[scale=0.06,angle=90,origin=c]{wavefunction2.png}}
%\end{picture}

\bibliography{apssamp}

\appendix

\end{document}

\title{SUPPLEMENTARY MATERIAL \\ ``Rydberg Entangling Gates in Silicon''}% 
\author{E. Crane}
\affiliation{Department of Electrical Engineering and London Centre for Nanotechnology, University College London, Gower Street, London WC1E 6BT, United Kingdom}
\author{A. Schuckert}
\affiliation{Department of Physics, Technical University of Munich, 85748 Garching, Germany}
\author{N. H. Le}
\affiliation{Advanced Technology Institute and Department of Physics, University of Surrey, Guildford GU2 7XH, United Kingdom}
\author{A. J. Fisher}
\affiliation{Department of Physics and Astronomy and London Centre for Nanotechnology, University College London, Gower Street, London WC1E 6BT, United Kingdom}
\date{\today}

\maketitle
\tableofcontents
\twocolumngrid 

\section{Rydberg gates in the presence of decoherence}\label{ann_fid}
First, we present the Lindblad master equation approach to simulating the whole pulse sequence on two qubits in the presence of dephasing and spontaneous decay. Second, we present the original resonant gate and its optimisation~\cite{Jaksch2000}. Third, we present the recently proposed ultrafast off-resonant blockade gate~\cite{levine_parallel_2019} discussed in the main text. Finally, we present our gate which shows much higher fidelities than the two previous proposals.  

To benchmark the gates, we calculate the fidelity of creating the Bell-state $|\Phi^+\rangle=\frac{1}{\sqrt{2}}|00\rangle+|11\rangle$ by simulating the whole pulse sequence in the presence of decoherence. We model a two-donor system with Hamiltonian~\cite{Jaksch2000} 
\begin{align}
H&=\sum_{i=1,2} \left(\frac{\Omega_i}{2} |1\rangle_i\langle r|+\frac{\Omega_i^*}{2} |r\rangle_i\langle 1|-\Delta|r\rangle_i\langle r|\right)\notag\\&+u\,|r\rangle_1\langle r|\otimes |r\rangle_2\langle r|,
\end{align}
where $\Omega_i,\Delta,u$ are Rabi frequency of the laser acting on donor $i$, detuning between the laser frequency and the transition frequency between $\ket{1}$ and $\ket{r}$, and interaction strength between the two donors, respectively. We restrict the donor energy levels to the two qubit levels $\ket{0},\ket{1}$ and the Rydberg state $\ket{r}$. In order to take into account dephasing and spontaneous emission from the short-lived Rydberg state, we solve a Markovian Lindblad Master equation for the density matrix $\rho$ of the form
\begin{equation}
    \partial_t \rho = -\frac{i}{\hbar} \left[H_{\mathrm{eff}}\rho-\rho H_{\mathrm{eff}}^\dagger\right]+\sum_j L_j \rho L_j^\dagger,
    \label{eq:Lindblad}
\end{equation}
where $H_{\mathrm{eff}}=H-\frac{i}{2}\sum_j L_j^\dagger L_j$ and $L_j$ are the jump operators given by
\begin{align}
L_{\mathrm{de},1}&=\sqrt{\gamma_\mathrm{de}}\left[(\ket{r}\bra{r}-\ket{1}\bra{1})\otimes \mathds{1}\right],\\
L_{\mathrm{de},2}&=\sqrt{\gamma_\mathrm{de}}\left[\mathds{1}\otimes(\ket{r}\bra{r}-\ket{1}\bra{1})\right],\\
L_{\mathrm{se},1}&=\sqrt{\gamma_\mathrm{se}}\left[\ket{1}\bra{r}\otimes\mathds{1}\right],\\
L_{\mathrm{se},2}&=\sqrt{\gamma_\mathrm{se}}\left[\mathds{1}\otimes\ket{1}\bra{r}\right],
\end{align}
where $\gamma_\mathrm{se},\gamma_\mathrm{de}$ denote the rates of spontaneous emission of phonons and dephasing between the excited state $\ket{r}$ and the qubit level $\ket{1}$. We assumed spontaneous emission does not bring $\ket{r}$ to $\ket{0}$ and we neglect decoherence processes between the qubit levels as these are negligible on the time scale of a single two-qubit gate.

We solve the evolution for arbitrary decoherence rates and interactions and express all results in units of the decoherence rate $\gamma_\mathrm{se}$, assuming that $\gamma_\mathrm{de}=0.5 \gamma_\mathrm{se}$. In order to deduce the necessary interaction strength and Rabi frequency for a successful gate operation, the corresponding value for $\gamma_\mathrm{se}$ given the donor/atom species has to be inserted as discussed in the main text.
%, realizing the operator
%begin{equation}
%-1\times\bordermatrix{&\ket{11}&\ket{10}&\ket{01}&\ket{00}\cr
%    \bra{11}&-1&0&0&0\cr
%    \bra{10}&0&-1&0&0\cr
%    \bra{01}&0&0&-1&0\cr
%    \bra{00}&0&0&0&1\cr}\notag.
%\end{equation}

We start in the initial state $|\Psi_0\rangle= (|1\rangle+|0\rangle)/\sqrt{2} \otimes (|1\rangle+|0\rangle)/\sqrt{2}$, and entangle the two qubits by numerically solving the time evolution of the pulse sequence. Finally, we apply a single qubit phase gate to rotate to $\ket{\Phi^+}$~\cite{levine_parallel_2019}. We assume the latter to be perfect, i.e. we act with the appropriate unitary operator on the density matrix instead of simulating the pulse. We use the fidelity of creating $|\Phi^+\rangle$, defined as
\begin{equation}
\mathcal{F}=\bra{\Phi^+}\rho\ket{\Phi^+}
\end{equation}
 as a measure how much decoherence alters the success of the gate.

\subsection{Resonant blockade gate~\cite{Jaksch2000}}

\begin{figure}[H]
\centering
    \begin{tabular}{c|cccc}

Pulse sequence & $\ket{11}$ & $\ket{10}$ & $\ket{01}$ & $\ket{00}$ \\ 
\hline 
1. $\pi$-pulse on atom 1 & i$\ket{r1}$ & i$\ket{r0}$ & $\ket{01}$ & $\ket{00}$ \\ 

2. $2\pi$-pulse on atom 2 & i$\ket{r1}$ & i$\ket{r0}$ & i$\ket{0 r}$ & $\ket{00}$ \\ 

3. $\pi$-pulse on atom 1 & -$\ket{11}$ & -$\ket{10}$ & -$\ket{01}$ & $\ket{00}$ \\ 

\end{tabular}
\includegraphics{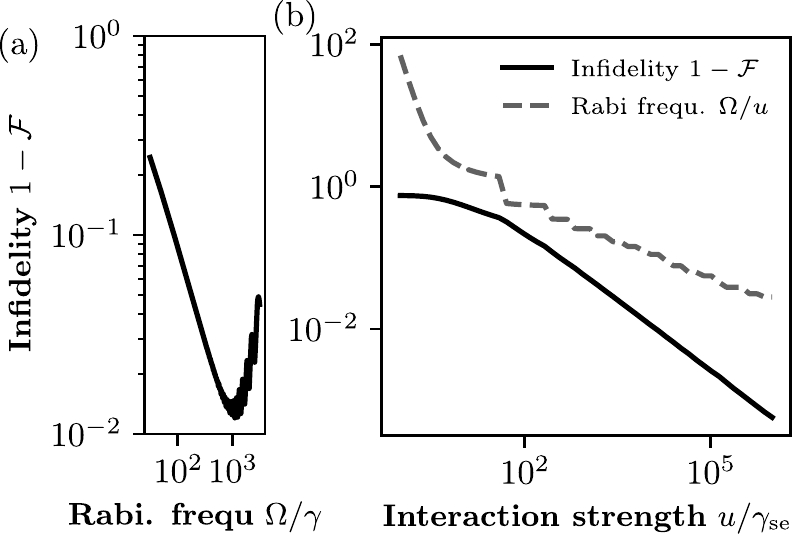}
\caption{\textbf{Fidelity of the resonant Rydberg blockade gate.--} Table: summary of the pulse sequence~\cite{Jaksch2000} in the ideal case of $u/\Omega=\infty$ and $u/\gamma=\infty$, showing the implementation of the controlled-Z truth table. (a) Optimal Bell-state $\ket{\Phi^+}$ fidelity obtained from a Lindblad simulation of the pulses, as a function of Rabi frequency for a fixed interaction strength $u=10^4\gamma_\mathrm{se}$: there is a clear optimal value. (b) Optimal fidelities as a function of interaction strength, showing that modest fidelities can be already reached for small  values of the interaction strength.\label{fig_Jaksch}}
\end{figure}

The resonant Rydberg blockade gate requires the qubits to be individually addressable and the laser to be on resonance with the orbital transition. %  such that in the general wavefunction $\ket{\psi(t)} = c_1(t)e^{-i E_1 t/\hbar}\ket{\psi_1}+c_2(t)e^{-i E_2 t/\hbar}\ket{\psi_2}$ which depends on the duration of the interaction $t$, the factors $c_1(t)=\cos{\frac{\Omega t}{2}}$ and $c_2(t)=-i\sin{\frac{\Omega t}{2}}$, proportional to the probability of being in the ground $\ket{\psi_1}$ or excited $\ket{\psi_2}$ state, are equal to -1 and 0 respectively for a 2$\pi$ pulse (t=2$\pi/\Omega$).
The pulse sequence (c.f. Fig.\ref{fig_Jaksch}) applies a $\pi$ pulse to the first atom, a $2\pi$ pulse to the second atom and a $\pi$ pulse to the first atom again. In the initial state $\ket{00}$ the pulse sequence has no effect on the qubits because the $\ket{0}$ state is off-resonant with the laser. Due to the Rydberg blockade, $\ket{11}$ acquires the same phase as $\ket{10}$, as the second atom cannot be excited to $\ket{r}$, in total implementing the truth table of a controlled-Z gate. %However, for finite interaction strength $u$, the second atom has a finite probability to get excited to $\ket{r}$, introducing a small phase $\phi\approx \pi \Omega / 2 u \ll \pi$.

%We follow the following sequence, starting from the initial state $\ket{\Psi_0}=\frac{1}{2}\left(\ket{0}+\ket{1}\right)\otimes\left(\ket{0}+\ket{1}\right)$ and on resonance, $\Delta=0$:
%\begin{enumerate}
%    \item \textbf{$\pi$-pulse on donor $1$.--} Evolve for time $\pi/\Omega$ with $\Omega_1=\Omega, \Omega_2=0$.
%    \item \textbf{$2\pi$-pulse on donor $2$.--} Evolve for time $2\pi/\Omega$ with $\Omega_1=0, \Omega_2=\Omega$.
%    \item \textbf{$\pi$-pulse on donor $1$.--} Evolve for time $\pi/\Omega$ with $\Omega_1=\Omega, \Omega_2=0$.
%    \item \textbf{Single qubit phase gate.--} Apply (decoherence free) single qubit phase gate $\left[\ket{0}\bra{0}+\exp(-i\phi)\ket{1}\bra{1}\right]$ on both donors.
%\end{enumerate}
In the presence of a non-zero decoherence rate $\gamma$, there is an optimum for the Rabi frequency $\Omega$: If $\Omega\approx\gamma$, the fidelity is low because the pulses take too long, however if $\Omega\approx u$, the fidelity is also low as the blockade condition is not well fulfilled.  We hence optimize the Rabi frequency $\Omega$ to maximize the fidelity $\mathcal{F}$.

In Fig.~\ref{fig_Jaksch} we show that this intuitive picture is correct, i.e. there is a clear optimum value for the Rabi frequency. Interestingly, the optimal Rabi frequencies are rather large showing that the gate operation time should be as small as possible to reduce loss from the excited state. This means that the hierarchy of scales is given by $\gamma \ll \Omega < u$. Importantly for the purposes of the donor implementation, we also find that for a given value of the Rabi frequency, the fidelity is only weakly dependent on the exact value of $u$. For large $u$, a plateau is reached beyond which the fidelity does not increase due to decoherence processes then being the limiting factor. At this point, the Rabi frequency should be increased to increase fidelities further.

\subsection{Off-resonant blockade gate~\cite{levine_parallel_2019}}

\begin{figure}[H]
\centering
\begin{picture}(250,220)
\put(20,150){\includegraphics[scale=0.3]{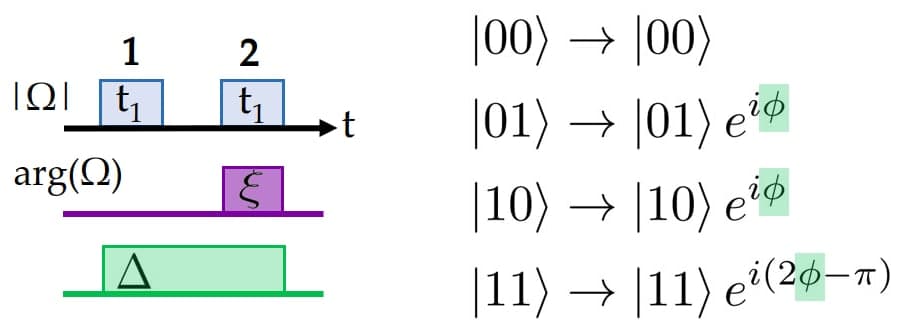}}
\put(30,30){\includegraphics[scale=0.17]{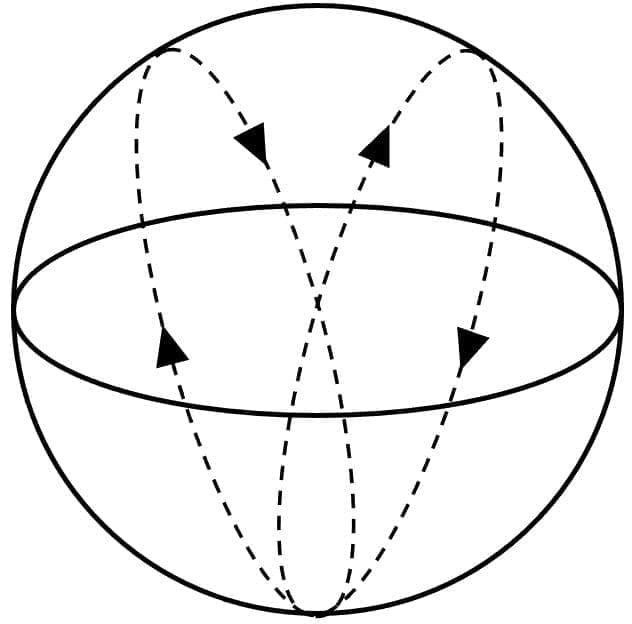}\quad\qquad \includegraphics[scale=0.2]{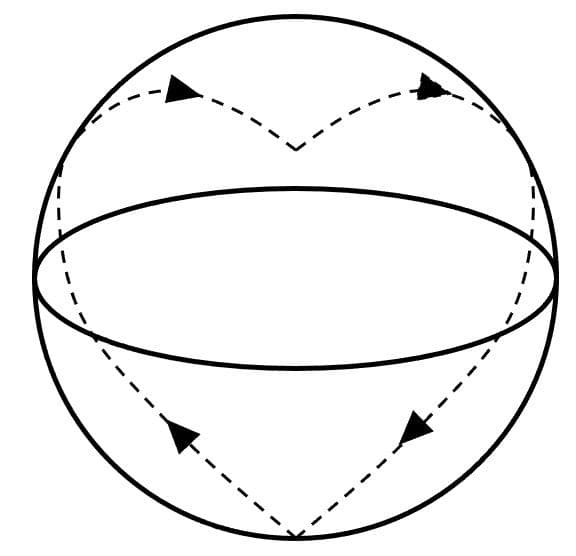}}
\put(0,220){a)}
\put(110,220){b)}
\put(0,125){c)}
\put(102,100){$\sqrt{2}\Omega$ \textbf{RB}}
\put(222,100){$\Omega$}
\put(177,115){$\ket{0r}$}
\put(177,20){$\ket{01}$}
\put(50,115){$\ket{1r}+\ket{r1}$}
\put(65,20){$\ket{11}$}
\put(0,-10){d)}
\end{picture}
\includegraphics[scale=0.95]{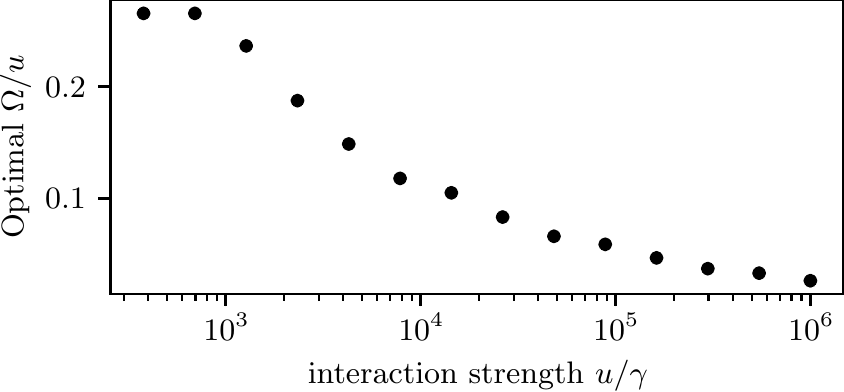}
\caption{\label{fig_levine}\textbf{Pulse sequence of the off-resonant gate} Adapted from Ref.~\cite{levine_parallel_2019}. a) Global pulse sequence acting with $|\Omega|$ on both qubits simultaneously: $t_1$ pulse duration chosen such that the first pulse returns $\ket{11}$ to itself (whereas $\ket{01}$ and $\ket{10}$ are left in an arbitrary location on the Bloch sphere), $\xi$ phase of the second pulse (evolve with $\Omega\exp(i\xi)$) chosen such that $\ket{01}$ and $\ket{10}$ return to themselves after the pulse sequence, $\Delta$ detuning of the $\ket{1}$ to $\ket{r}$ transition chosen such that the phases highlighted in b) are equal. Finally, the third step is to apply single qubit phase gates to both qubits, with phase $\phi$, which corrects for global single qubit phases built up in the dynamics. b) Mapping of the qubit states due to the Cz gate. c) The dynamics of the states $\ket{11}$ (with Rabi frequency $\sqrt{2}\Omega$ in the case of a perfect blockade) and $\ket{0 1}$ (with Rabi frequency $\Omega$, leading to a different path over the Bloch sphere) in terms of two level systems. d) Optimal Rabi frequency in the off-resonant blockade gate as a function of interaction strength and decoherence time.}
\end{figure}

In the improved blockade gate as presented and implemented in Ref.~\cite{levine_parallel_2019}, only two global pulses with fixed detuning $\Delta\neq 0$ are needed. The pulse sequence proceeds as follows:
\begin{enumerate}
    \item Evolve for time $\tau$ with $\Omega_1=\Omega_2=\Omega$. 
    \item Evolve for time $\tau$ with $\Omega_1=\Omega_2=\Omega\exp(i\xi)$,
\end{enumerate}
again followed by a single qubit phase gate on both qubits (c.f. Fig.~\ref{fig_levine}a)). In the above sequence, the gate time $\tau$ is chosen such that the state $\ket{11}$ returns to itself after the first pulse, i.e. to the time corresponding to a blockaded $2\pi$ pulse, $\tau=2\pi/\sqrt{2\Omega^2+\Delta^2}$. Similarly, the phase $\xi$ of the laser in the second pulse is chosen such that the state $\ket{01}$ returns to itself after both pulses. Finally, the detuning $\Delta$ is chosen such that the phase acquired by both states differs by $\pi$ (in the sense that this phase difference remains after application of the global single qubit phase gate, c.f. Fig.~\ref{fig_levine}b).

In the strongly interacting regime, the parameters were analytically calculated to be given by~\cite{levine_parallel_2019}
\begin{align}
\Delta/\Omega&=0.377371\\
\xi&=3.90242.
\end{align}
This leaves the Rabi frequency as a free parameter, which we optimize for a given interaction strength $u$ and decoherence rates $\gamma_\mathrm{se}$ and $\gamma_\mathrm{de}=0.5\gamma_\mathrm{se}$. Similarly to the resonant gate, we find a clear optimal Rabi frequency for a given interaction strength, which we show in Fig.~\ref{fig_levine} and find slightly higher fidelities than for the resonant gate (see main text).

A major limitation for both blockade gate proposals is the hierarchy of scales $\gamma\ll \Omega < u$, given by the requirement for the blockade condition to be fulfilled. This limits the maximum Rabi frequency, setting a lower bound for the gate duration and leaving more time for decoherence processes to kick in. Our calculation implies that the experiment in Ref.~\cite{levine_parallel_2019} operated very close to the optimal Rabi frequency, in part explaining their improvements over previous results.

\subsection{Blockade-inspired off-resonant gate}

\begin{figure}
\centering
\begin{picture}(250,220)
\put(20,150){\includegraphics[scale=0.3]{pulseseq.png}}
\put(0,10){\includegraphics[scale=0.17]{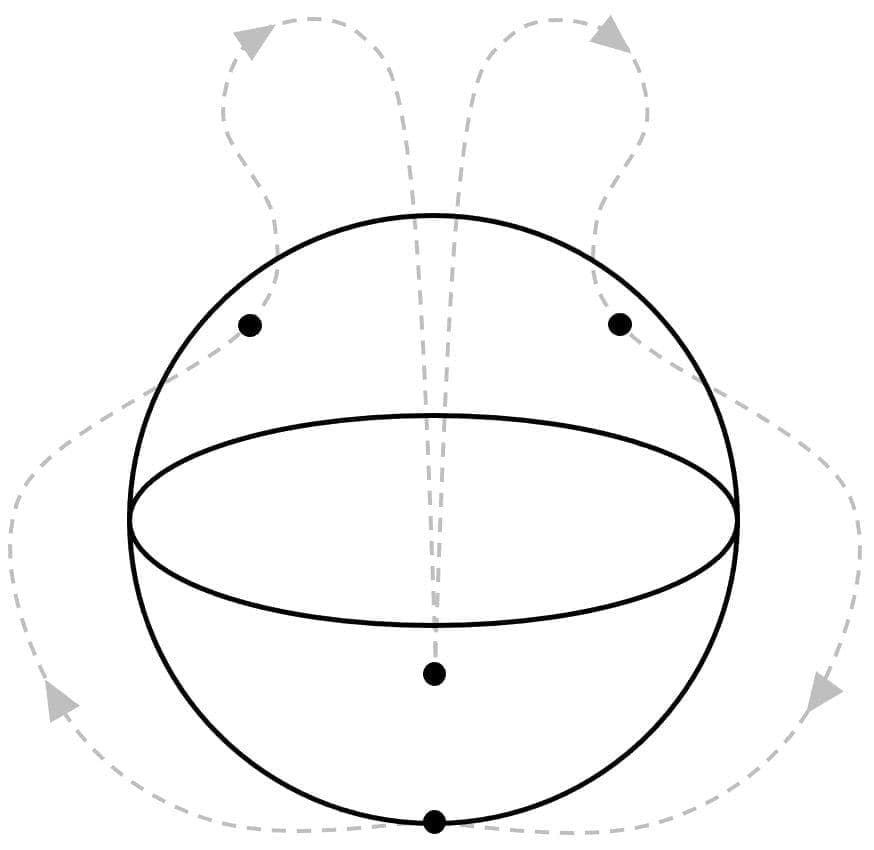}\quad\qquad \includegraphics[scale=0.17]{path2.png}}
\put(0,220){a)}
\put(110,220){b)}
\put(0,125){c)}
\put(85,120){\color{gray} $c_{11}(t)\ket{11}+c_{1r}(t)\ket{1r}$}
\put(90,105){\color{gray} $+c_{r1}(t)\ket{r1}$}
\put(85,90){\color{gray} $+c_{rr}(t)\ket{rr}$}
\put(177,95){$\ket{0r}$}
\put(177,0){$\ket{01}$}
\put(50,95){$\ket{rr}$}
\put(50,0){$\ket{11}$}
\put(0,-20){d)}
\put(0,-120){e)}
\end{picture}
\newline\newline\newline
\includegraphics{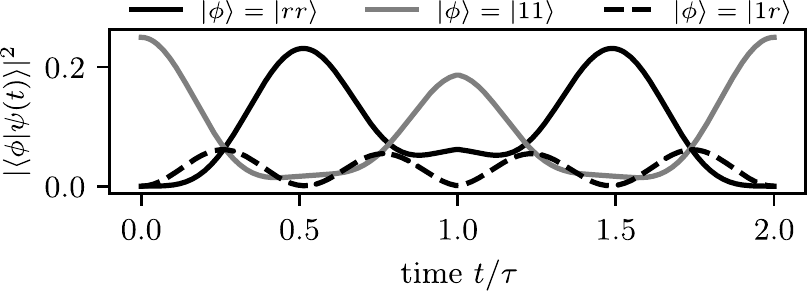}
\newline\newline
\includegraphics{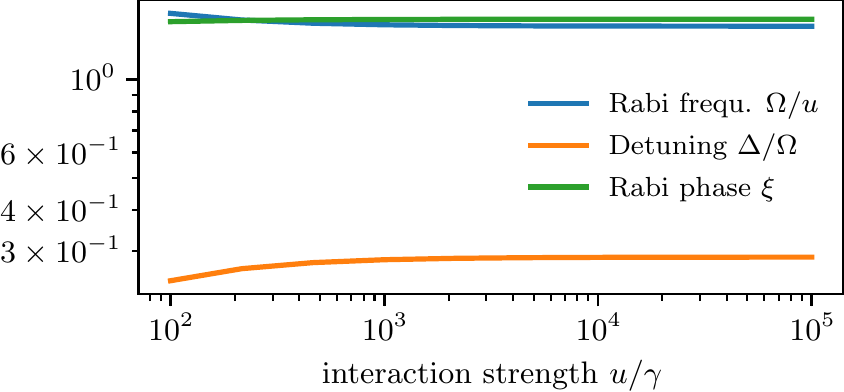}
\caption{\label{fig_new}\textbf{Pulse sequence of proposed gate} a) Global pulse sequence acting with $|\Omega|$ on both qubits simultaneously: pulse duration $\tau$ chosen such that the first pulse returns $\ket{01}$ to itself (whereas $\ket{11}$ is left in an arbitrary location on the Bloch sphere), $\xi$ is the phase of the second pulse chosen such that $\ket{11}$ returns to itself after the second pulse, $\Delta$ detuning of the $\ket{1}$ to $\ket{r}$ transition chosen such that the phases highlighted in b) are equal. c) Bloch-sphere depiction of the dynamics of the states $\ket{11}$ and $\ket{01}$. The $\ket{11}$ state touches the Bloch sphere spanned by $\ket{11}$ and $\ket{rr}$ at times $(n/2)\tau$ with $n=1,2,3,4$. This is more clearly shown in d), which depicts the state probabilities when starting in the initial state $\ket{\psi(0)}=(\ket{0}+\ket{1})\otimes (\ket{0}+\ket{1})/2$. e) Optimized pulse parameters in the blockade-inspired off-resonant gate. Note the very weak dependence on the interaction strength.}
\end{figure}

We propose a ``phase accumulation'' variant of the above gate, similar to the ``Gate A'' discussed in Jaksch et al.~\cite{Jaksch2000}.  To do so, we use the same pulse sequence as in the off-resonant blockade gate but reverse the reasoning - instead of tuning the gate time to the blockaded Rabi frequency, we use the unblockaded one, given by $\tau=2\pi/\sqrt{\Omega^2+\Delta^2}$. Hence, the states $\ket{01}$ and $\ket{10}$ return to themselves after a single Rabi pulse. Accordingly, we tune the phase $\xi$ such that the state $\ket{11}$ returns to itself after the second pulse, accumulating a two-qubit phase due to the non-vanishing probability to populate $\ket{rr}$. That this proposal is advantageous compared to the blockade gates may seem counterintuitive at first due to the high loss rate from $\ket{r}$. However, in this gate the condition $\Omega<u$ is relaxed and hence the gate duration can be much smaller compared to the blockade gates.  We optimize the pulse sequence and find that the optimal Rabi frequency is proportional to the interaction strength. Furthermore, all other parameters to be almost independent of the interaction strength. We obtain these results numerically by fixing $\tau=2\pi/\sqrt{\Omega^2+\Delta^2}$ and optimizing $\Omega,\Delta, \xi$ for a given $u/\gamma$, obtaining the optimal pulse parameters shown in Fig.~\ref{fig_new}. In the large interaction limit, we find
\begin{align}
\Omega/u&=1.45747,\\
\Delta/\Omega&=0.28757,\\
\xi&=1.5306.
\end{align}

Note that the optimal Rabi frequency is about an order of magnitude larger than in the blockade gate schemes (given a fixed value of the interaction strength), leading to an order of magnitude smaller gate durations and to the much larger fidelities shown in the main text.

%\begin{figure}
%\centering
%\includegraphics{2010_Fidelity.pdf}
%\caption{\label{fig_fidcomp}%\textbf{Fidelity of off-resonant (Levine) and resonant (Jaksch) blockade gates} as a function of interaction strength and decoherence time. Independent of the platform.}
%\end{figure}

%In the case of a sign change of the Rydberg blockade energy $u$, as for example present for the 1sT$_2\Gamma_7$ state of Si:Se+, it is possible to flip the sign of the detuning in order to obtain the same fidelities.

\subsubsection{Robustness to placement error}
\begin{figure}[H]
\includegraphics{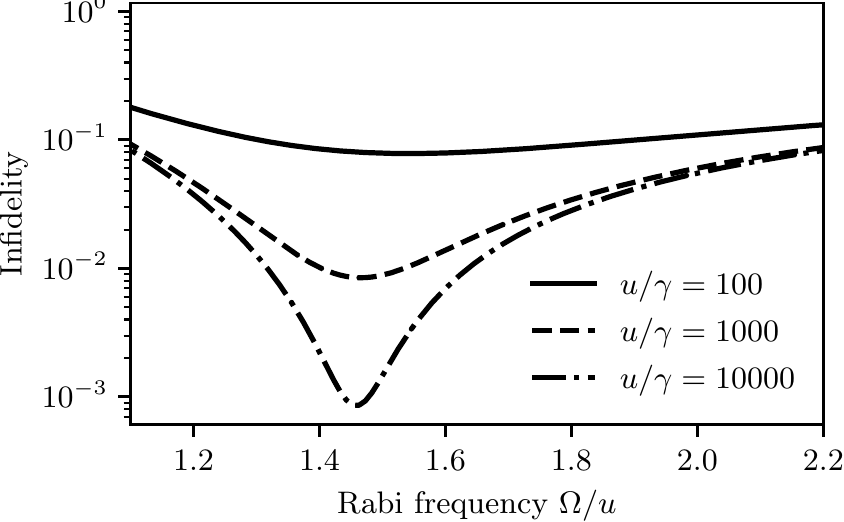}
\caption{\label{fig_fid_res}\textbf{Dependence of the fidelity on the Rabi frequency} for the blockade-inspired off-resonant gate for some examples of the interaction strength.} 
\end{figure}
We checked that a displacement within the 2x3 lattice site physical limit on the precision placement of hydrogen lithography of phosphorous in silicon~\cite{Fuechsle2012} would still be within the width of the ``fidelity resonance'', as would be a small diffusion out of plane, see Fig.~\ref{fig_fid_res}. This result implies that a sample made with hydrogen ligthography would not need tuning - assuming the interaction between donors as a function of distance is known, and for an ion implaneted sample, a simple procedure to finding good pulse parameters could be a rough scan of the sample with an scanning tunnelling microscope (STM) after fabrication to determine the locations of the donors. Note that while placement-insensitivity is stronger in the blockade gates, there is still an error introduced by $\Omega/u$ differing from the optimal value if $u$ varies.
\subsubsection{Robustness to inhomogeneous broadening}
\begin{figure}[H]
\centering
\includegraphics[scale=1]{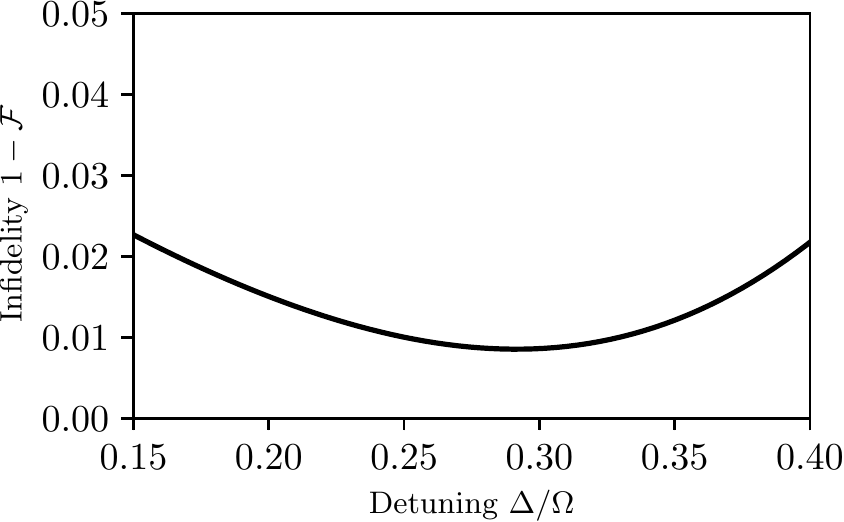}
\caption{\label{fig_detunings} \textbf{Negligible effect of inhomogeneous broadening.--} Infidelity as a function of interaction strength scaled by $\ket{r}$ lifetime, for a fixed $\Omega$ and interaction strength $u$. Even a relatively large inhomogeneous broadening of $0.1\Omega$ corresponding to roughly $0.2$meV for Si:P, which is a factor of two larger than found in experiment~\cite{Litvinenko2015}, has very little effect on the fidelity.}
\end{figure}
Inhomogeneous broadening refers to a change in the excited state energies due to a differing electrostatic environment of the donor, e.g. caused by randomly distributed silicon isotopes. For the Si:P 1s-2p+- transition, this effect can result in a variation of excited state energies by about $0.1$meV as measured in Ref.~\cite{Litvinenko2015}. In our gate scheme, not taking into account this effect leads to an error in the detuning chosen for the pulse. In order check that this leads to insubstantial changes in the gate fidelity, we plot the infidelity for a large range of detunings around the optimal value in Fig.~\ref{fig_detunings} for a fixed Rabi frequency $\Omega=1000\gamma$ (corresponding to $2.8 meV$ for the Si:P lifetime, such that a variation of $0.1$ meV in the detuning corresponds to about $0.04\Omega$). The fidelities are not heavily affected, and we conclude that the gate does not need tuning to account for inhomogeneous broadening.
%\begin{figure}
%\centering
%\includegraphics[scale=0.92]{block_fid-tdec.pdf}\quad
%\raisebox{10mm}[0pt][0pt]{\includegraphics[scale=0.32]{legf.png}}
%\caption{\label{fig_fidexp} Infidelity as a function of interaction strength for various $\ket{r}$ lifetimes and optimised pulse parameters (platform-independent results). Experimental: Se+ 1sT$_2\Gamma_7$ 7.7ns ~\cite{DeAbreu2019} and P 2p0 235ps~\cite{Hubers2013}. Theoretical: P 2p0 1ns ~\cite{Tyuterev_Theoretical_2010}.}
%\end{figure}

\section{Interactions}\label{ann_ints}

\begin{figure*}
\centering
\begin{picture}(445,220)
\put(-32,170){Se+}
\put(-32,160){2p0}
\put(-32,70){Se+}
\put(-32,60){1sT2}
\put(-10,100){\includegraphics[scale=0.95]{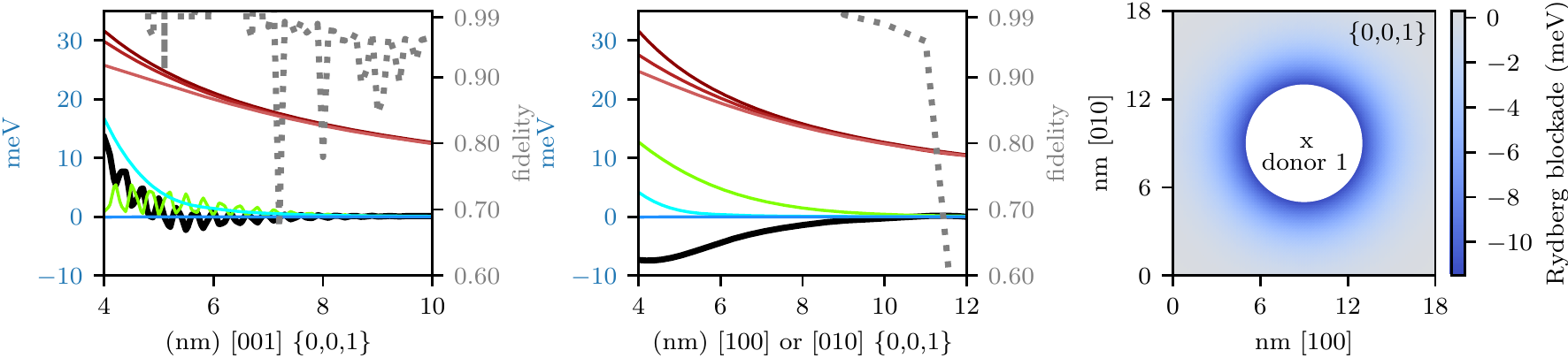}}
\put(-10,0){\includegraphics[scale=0.95]{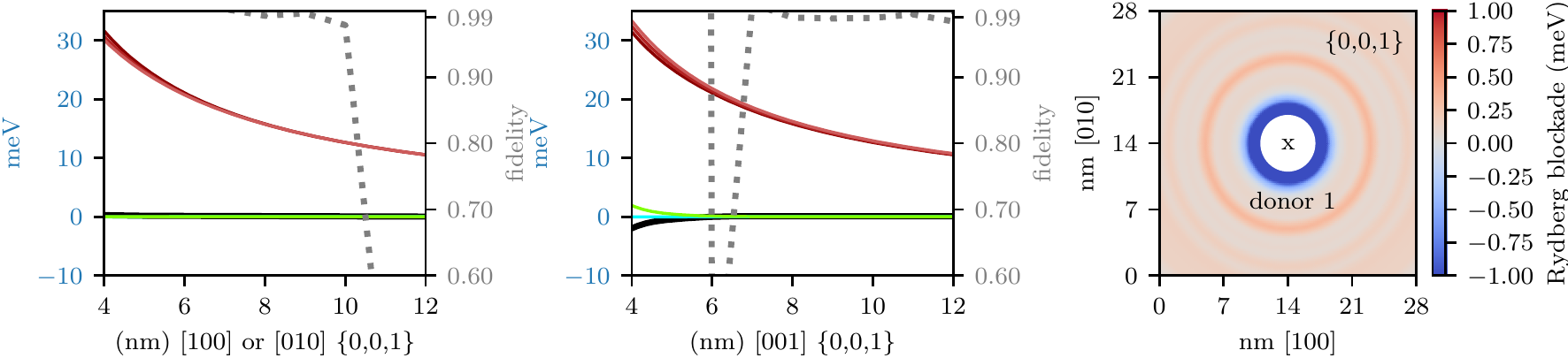}}
\put(-10,200){a)}
\put(160,200){b)}
\put(320,200){c)}
\put(-10,100){d)}
\put(160,100){e)}
\put(320,100){f)}
\end{picture}
\caption{\label{fig_ann_Se+_ints} \textbf{Rydberg blockade between \textsuperscript{77}Se\textsuperscript{+} excited states} with polarisation \{0,0,1\}. Row 1: interactions between two 2p0 states and fidelity assuming a lifetime of 1ns. Row 2: interactions between two 1sT$_2\Gamma_7$ states and fidelity for the experimentally determined lifetime 7.7ns~\cite{Morse2017}. Color labelling identical to Fig. 2 in main text. Reds '$W$': inter-site Coulomb electron - electron repulsion ('$rr$': 2p0-2p0, '$rg$': 2p0-1sA and '$gg$': 1sA-1sA). Green, '$J$': ferromagnetic exchange (2p0-2p0). Light blue, '$V_{V\!dW}$': Van der Waals (2p0-2p0). Dark blue, '$V_{V\!dd}$': induced dipole interaction with non-ionising field (negligible in all cases shown above), '$u$': Total interaction calculated from Eq.~\ref{eq_ryd}. (a and d) Inter-donor axis parallel to polarisation axis. (b and e) Inter-donor axis perpendicular to polarisation axis. (c and f) Map of total interaction (meV) for one donor placed at the center of the map and the other to occupy any other position. White: avoiding the molecular limit. a) Oscillation due to 'J'. e) Fidelity dips to 0 at 6nm because $u$ changes sign, for which one can change the sign of the detuning. Fidelity remains large due to a difference on the order of $0.1$ meV coming from the 'W' terms and the extremely long decay time of the 1sT$_2\Gamma_7$ state.}
\end{figure*}

With regards to the interactions between donors, we need to consider the tight binding model, with additional dipolar forces due to a donor being made up of a core plus a valence electron.
\begin{equation}
H = \sum_{i\neq j}a^{\dagger}_{i\sigma}t_{ij}a_{j\sigma} + \sum_{ii'jj'} \frac{U_{ii'jj'}}{2}a^{\dagger}_{i\sigma}a^{\dagger}_{i'\sigma'}a_{j'\sigma'}a_{j\sigma}
\end{equation}
where t is the single hopping matrix element between neighboring sites and
\begin{align}
U_{ii'jj'}=\int d^d\bm{r}_1 d^d\bm{r}_2 \, &\psi_{R_i}^{*}(\bm{r}_1)\psi_{R_i'}^*(\bm{r}_2)V(\bm{r}_1-\bm{r}_2)\notag \\ &\times\psi_{R_j}(\bm{r}_1)\psi_{R_j'}(\bm{r}_2),
\end{align}
where V is the Coulomb interaction, given by $V(\bm{r})=V_0/\bm{r}$ where $V_0=\frac{e^2}{4 \pi \epsilon_0 \epsilon_S}$, $e$ is the electron charge and $\epsilon_S=11.4$ is the dielectric constant of silicon. 

The contribution from $U_{ii'ii'}=W_{ii'}$ is $\sum_{i\neq i'}W_{ii'}\hat{n_i}\hat{n_{i'}}$ where $\hat{n_i}=\sum_{\sigma}a^\dagger_{i \sigma}a_{i \sigma}$ which corresponds to the essentially classical Coulomb interactions between donors on neighbouring sites. From this we get:
\begin{itemize}
\item \textbf{Intersite Coulomb electron - electron repulsion}
	\begin{equation}
	W_{12}=V_0\int\int d\bm{r}_1 d\bm{r}_2 \frac{|\psi_1 (\bm{r}_1)|^2 |\psi_2 (\bm{r}_2-\bm{R})|^2 }{|\bm{r}_1 - \bm{r}_2|},
	\end{equation}
where $\bm{R}=\bm{R}_2-\bm{R}_1$ is the separation vector between the two donors. 
\end{itemize}

The contribution from $U_{iiii}$ corresponds to the on-site interaction, which can lead to anti-ferromagnetic super-exchange which is negligible in our case.

The contributions from $U_{ijji}$ can be re-written, making use of Pauli matrix identities, into:
\begin{equation}
\sum_{i \neq j}\frac{U_{ijji}}{2} a^\dagger_{i \sigma}a^\dagger_{j \sigma'}a_{i \sigma'}a_{j \sigma} = - \sum_{i \neq j} U_{ijji}(\overrightarrow{S_i}\cdot\overrightarrow{S_j}+\frac{1}{4}\hat{n_i}\hat{n_j}).
\end{equation}
It corresponds to the ferromagnetic exchange coupling. Note that for calculating the total interaction for the gate, we are only interested in the case where both spins are on resonance with the laser, so both spins are aligned. This leads to $\overrightarrow{S_i}\cdot\overrightarrow{S_j}=\frac{1}{4}\hat{n_i}\hat{n_j}$.
\begin{itemize}
\item \textbf{Ferromagnetic exchange coupling}
We refer to $U_{ijji}$ as $J_{ij}$ and we have:
\begin{align}\label{eq:HL}
J_{12}=\int d\bm{r}_1 d\bm{r}_2 \, &\psi_1^{*}(\bm{r}_1)\psi_2^{*}(\bm{r}_2-\bm{R})\frac{V_0}{|\bm{r}_1-\bm{r}_2|} \notag \\&\times\psi_2(\bm{r}_1-\bm{R})\psi_1(\bm{r}_2),
\end{align}
where $\bm{R}=\bm{R}_2-\bm{R}_1$ is the separation vector between the two donors.
\end{itemize}
The leading correction to this electron-only model of donor interactions is given by dipole interactions.
\begin{itemize}
\item \textbf{Dipole-dipole interactions} take the form:
\begin{align}
V_{V\!dd} &= \frac{\bm{p}_1\bm{p}_2 - 3(\bm{n}\cdot\bm{p}_1)(\bm{n}\cdot\bm{p}_2)}{R^3},
\end{align}
where $\bm{p_i}$ is the dipole moment of atom $i$, $R$ is the inter-atomic distance, and $\bm{n}$ is the unit vector between the two donors.
\end{itemize}

$V_{V\!dd}$ is the dominant interaction in cold Rydberg atoms. At first order, the dipole-dipole interaction cancels, due to the vanishing dipole moment of donors. However, when an electric field is applied, dipole moments $\bm{p_i}$ are induced and can be calculated as the derivative of the Stark shift of the state $i$ which we study in App. \ref{ann_Stark}.

A dipole can furthermore be induced by the vacuum, leading to Van der Waals forces, which are calculated in second order perturbation theory of $V_{V\!dd}$.
\begin{itemize}
\item \textbf{Van der Waals} forces:
\begin{equation}
\begin{split}
& \qquad \qquad \qquad \quad V_{V\!dWij} = - \mathop{\sum_{n',l',j',m'}}_{n'',l'',j'',m}\\ &\frac{|\bra{n',l',j',m'} \otimes\bra{n'',l'',j'',m''}V_{V\!dd}\ket{n,l,j,m}\otimes\ket{n,l,j,m}|^2}{E_{\ket{n',l',j',m'}}+E_{\ket{n'',l'',j'',m''}}-2E_{\ket{n,l,j,m}}}
\end{split}
\end{equation}
\end{itemize}

To conclude, the Hamiltonian we use in this problem, in which we have checked that we can neglect the tunnelling at the distances we consider and we do not include electron-core interactions~\cite{Wu_Exchange_2008,Heitler_Interactions_1927} as they cancel in the expression of the total interaction in the main text, corresponds to:
\begin{align}
H = &\sum_{i\neq j}\frac{1}{2}(W_{ij}+V_{V\!dWij}-J_{ij})\hat{n_i}\hat{n_{j}}.\notag
\end{align}
In the presence of an electric field, $W$ is dropped in favor of $V_{V\!dd}$ which becomes the leading contribution. 

In the case of the entangling gate, we are interested in the energy difference occurring when both donors are excited to the Rydberg state, compared to when none or only a single of the two is excited, as mentioned in the main text, which corresponds to $E_{\ket{rr}}$, in which both spins are excited, $E_{\ket{rg}}$ in which one donor is excited and the other is in the ground state and $E_{\ket{gg}}$ in which both donors are in the ground state:
\begin{align}
E_{\ket{rr}} &= W_{\ket{rr}}+V_{V\!dW\ket{rr}}-J_{\ket{rr}}\\
E_{\ket{rg}} &= W_{\ket{rg}}\\
E_{\ket{gg}} &= W_{\ket{gg}}
\end{align}
where, again, $V_{V\!dd}$ dominates $W$ in the presence of an external electric field and we neglected the extremely small contributions coming from $V_{V\!dW}$ and $J$ for $\ket{rg}$ and $\ket{gg}$.

From these expressions, we are able to obtain the expression for the total interaction:
\begin{align}
u&=(E_{\ket{rr}}-E_{\ket{gr}})-(E_{\ket{rg}}-E_{\ket{gg}})\\
&\approx W_{\ket{rr}}-2W_{\ket{rg}}+W_{\ket{gg}}-J_{\ket{rr}}+V_{V\!dW\ket{rr}}\label{eq_ryd},
\end{align}
$J_{\ket{gg}}, V_{V\!dW\ket{gg}}, V_{V\!dW\ket{rg}},J_{\ket{rg}}$ are negligible for the range of donor separations considered in this work.
The interaction energy is plotted as a map for the 2p0 and 1sT$_2\Gamma_7$ states in Se+ in Fig.~\ref{fig_ann_Se+_ints}.

\begin{figure}
\centering
\includegraphics[scale=1]{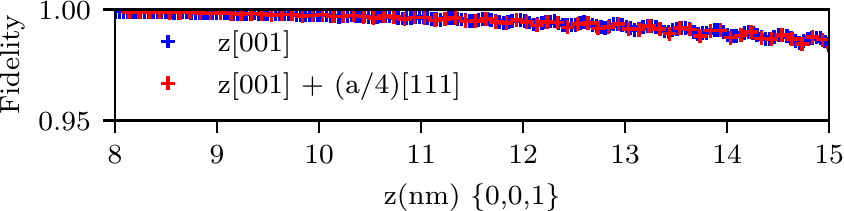}
\caption{\label{fig_offset} In the case of Si:P with $\ket{r}$ chosen as 2p0, as is the case in the main text, the fidelity is robust to donors belonging to different silicon sublattices.}
\end{figure}

For donor spin qubits in Si:P, most schemes make use of the ground state ferromagnetic exchange interactions, which oscillates wildly depending on the sublattice on which the donors reside. In Fig.~\ref{fig_offset} we show that in the gate proposed in this paper, the fidelity is not heavily affected by donor qubits residing on different sublattices. 

\section{Finite element method multivalley wavefunctions}\label{ann_FEM}

In the following we explain how to obtain multivalley wavefunctions of donor electrons in silicon, which result from coupling the hydrogenic wavefunctions (envelope functions) into the manifolds determined by the symmetry group of the system. The envelope wavefunctions can be obtained by solving the Schr\"odinger equation for hydrogen, with the appropriate effective mass. We do this using the Finite Element Method~\cite{le2019} (developed in engineering for numerically computing approximations to solutions to partial differential equations and recently applied to donors~\cite{le2019}).

The effective mass approximation states that close to a band extrema, such as the conduction band minimum, each electron can be described by a mean-field Hamiltonian (every electron experiences the same average periodic potential) which is that of a single free electron with a modified mass (the effective mass) in an impurity potential.

Silicon has a face centred cubic lattice which leads to its brouillin zone (its primitive cell in reciprocal space) being a truncated octohedron. The latter contains six square faces, the centers of which are called X points. The conduction band minima in silicon are positioned at the X-points and are all equivalent. Using the KP method, the effective mass at a band extrema - such as the conduction band minima - can be determined. The effective mass corresponds to the dispersion of the energy $E_k$ as a function of momentum $k$ ($E_k = \frac{\hbar^2 k^2}{2 m}$). In silicon, the conduction band minima are anisotropic: they are ellipsoids which lie along the axis linking the center of the brillouin zone ($\Gamma$ point) to the valley's X point. In valley specific coordinates, we set this to be the z axis. The effective mass along z is referred to as longitudinal $m^*_l = 0.191 m_e$, where as along x and y it is transverse $m^*_t = 0.916 m_e$, in reference to the conduction band minimum ellipsoid, where m is the mass of a free electron. The kinetic energy term in valley-specific coordinates is then:
\begin{equation}
-\frac{\hbar^2}{2}(\frac{\partial^2}{\partial x^2}+\frac{\partial^2}{\partial y^2}+\gamma\frac{\partial^2}{\partial z^2})
\end{equation}
where $\gamma = m_t / m_l$.

The impurity potential is the Coulomb potential of the proton, which at large distances can be approximated as a point charge at the nucleus, felt by the valence electron:
\begin{equation}
V_{imp}(r) = -\frac{\hbar}{4 \pi \epsilon_0 \epsilon_r}\frac{e^2}{r}
\end{equation}
where e is the electron charge. However, the valley degree of freedom in multivalley semiconductors allows the wavefunctions to have a non-negligible probability of being at the core where the Coulomb approximation breaks down. To account for this, we add another potential to the Hamiltonian, widely referred to as the central cell potential $U_{cc}(r)$, which takes the form of a delta function at the core and which we model using a Heavyside step function~\cite{PhysRevB.4.3468}. The value of the central cell potential is determined using a bisection algorithm by requiring the energy of the solution of the FEM method to match the experimentally determined energies for the multivalley states.

It is also possible to add to the Hamiltonian the confining potential of an external electric field applied along the real-space z axis (which corresponds to the z-axis in valley-specific coordinates), which is:
\begin{equation}
E_f(r)=e E_f z.
\end{equation}

The Schr\"odinger equation which we will simplify in the following, in the valley-specific coordinates where z is the valley axis, is:
\begin{equation}
\big[ -\frac{\hbar^2}{2 m*}\nabla ^2 + V_{imp}(r) + U_{cc}(r) + E_f(r) \big] \psi_i (r) = \epsilon_i \psi_i (r).\label{schr_eq}
\end{equation}
The wavefunction is written
\begin{equation}
\psi(r) = F_i(r)\phi(k_j,r)\alpha_i.
\end{equation}
It is composed of the envelope function $F_i(r)$, where the indices $i$ runs over the hydrogenic states, the Bloch wavefunction $\phi(k_j,r)=e^{-i k_j.r} u_{k_j}(r)$ which is the product of a plane wave and a lattice periodic function $ u_{k_j}(r)$, where j runs over the six valleys, and lastly a multivalley parameter $\alpha_i$ which couples the function into various manifolds according to the $T_d$ point group mentioned above. We use the finite element method to obtain $F_i(r)$ and couple it to the Bloch wavefunction and into the multivalley manifolds in a second step. As all the conduction band minima are equivalent, the solution to this Schr\"odinger equation suffices for all the manifolds representing the $T_d$ point group.

We write the Hamiltonian in the atomic units $a_B=4 \pi \epsilon_0 \epsilon_R/e^2 m_t$ and $E_H = e^4 m_t/(\hbar 4 \pi \epsilon_0 \epsilon_R)^2$:
\begin{align}
H = &\frac{- E_H a_B^2}{2} (\frac{\partial^2}{\partial x^2}+\frac{\partial^2}{\partial y^2}+\gamma\frac{\partial^2}{\partial z^2}) - \\ \notag
&\frac{E_H a_B}{\sqrt{x^2+y^2+z^2}}) + U_{cc}\delta(x,y,z)+ \frac{E_h}{a_B e} e Ef z
\end{align}

We make four steps to simplify the Hamiltonian to improve FEM convergence:
\begin{itemize}
\item Transform to an anisotropic frame where  $z'=\sqrt{z}$ to have a symmetrised Hamiltonian.
\item Transform to spherical polar coordinates (the kinetic energy has spherical symmetry so this is allowed).
\item Choose the wavefunction $F_{i,m}(r) = e^{i m \phi} f_{j,m,\mu}(r,\theta) =  e^{i m \phi} \frac{1}{r} Y_{j,m,\mu}(r,\theta) $ where m is the magnetic quantum number, to obtain a 2D Hamiltonian.
\item Transform the semi-infinite plane to a finite rectangle (tangent space) $r = r_0 \tan{\eta}$ with $\eta \in [0,\frac{\pi}{2}]$ and $r_0$ a scaling factor which is best chosen comparable to the wavefunction of interest~\cite{le2019} for a compressed Hamiltonian. $Y_{j,m,\mu}(r,\theta)$ becomes $y_{j,m,\mu}(\eta,\theta)$.
\end{itemize}

The boundary conditions are then that the wavefunction must decay at infinity, which corresponds to $\alpha \rightarrow 0$ and $\alpha \rightarrow \frac{\pi}{2}$:
\begin{equation}
y(0,\theta) = y(\pi/2,\theta) = 0.
\end{equation}
Additionally, for wavefunctions with parity and magnetic quantum numbers of opposite parity $y(\eta,\theta)=-y(\eta,\pi-\theta)$ therefore at the boundary:
\begin{equation}
y(\eta,\pi/2) = - y(\eta,\pi/2) = 0.
\end{equation}

The final symmetrised polar compressed 2D Hamiltonian reads:
\begin{align}
H =& -\frac{E_H a_B^2}{2 r_0}\big( \cos^2{\eta}\frac{\partial}{\partial \eta}\big(\frac{m^2 \cot^2{\eta}}{\sin^2{\theta}}\big)+\\&\frac{\cot^2{\eta}}{\sin{\theta}}\frac{\partial}{\partial\theta}\big(\sin{\theta}\frac{\partial}{\partial \theta}\big) \big) - \frac{E_H a_B r_0^{-1} \cot{\eta}}{\sqrt{1-(1-\gamma) \cos^2{\theta}}} + \notag \\  &\frac{U_{cc}}{\frac{4}{3}\pi r_c^3} H(r-r_c) + \sqrt{\gamma} r \cos{\theta} E_f Us(r-r_m)
\end{align}
Where H(r) is the Heavyside Theta function and Us(r) is the Unitstep function and $r_c$ and $r_m$ cutoff values. The solutions obtained via FEM are then numerically normalised using the Cuba Vegas package~\cite{HAHN200578}.

We now describe the coupling to the multivalley manifolds. Firstly, we reduce the six valley problem $(-x,x,-y,y,-z,z)$ to an effective three valley problem $(x,y,z)$ with x,y and z $\in [-\infty,\infty]$. We associate to each valley the envelope function in rotated cartesian coordinates to ensure consistency when all the valleys will be coupled, considering we took z as the valley axis we create a three part vector corresponding to the envelope wavefunction in the three effective valleys: $(F_i(z,x,y),F_i(y,z,x),F_i(x,y,z))$.

We transform the six functions of the tetrahedral point group into the effective three valleys by taking the Bloch wavefunction into account and using trigonometric identities to couple two valleys of opposite sign, each with a plane wave factor $e^{-i k_i x}$ into one with $\sin{2k_i x}$ if the manifolds have an even total sign or $\cos{2 k_i x}$ if it is odd. For the 1sT$_2\Gamma_7$ state for example:
\begin{align}
&F_i(x)e^{-i k_i x} . (1,-1,0,0,0,0)\notag  \\ \notag = & F_i(x)e^{-i k_i x} - F_i(x)e^{-i k_i x} \\ \notag  = &2i \sin{2 k_i x} F_i(x).
\end{align}
We then normalise the wavefunctions.

\section{Ionisation rates and Stark shifts}\label{ann_Stark}

\begin{figure}
\centering
\begin{picture}(210,250)
\put(-10,0){\includegraphics[scale=1]{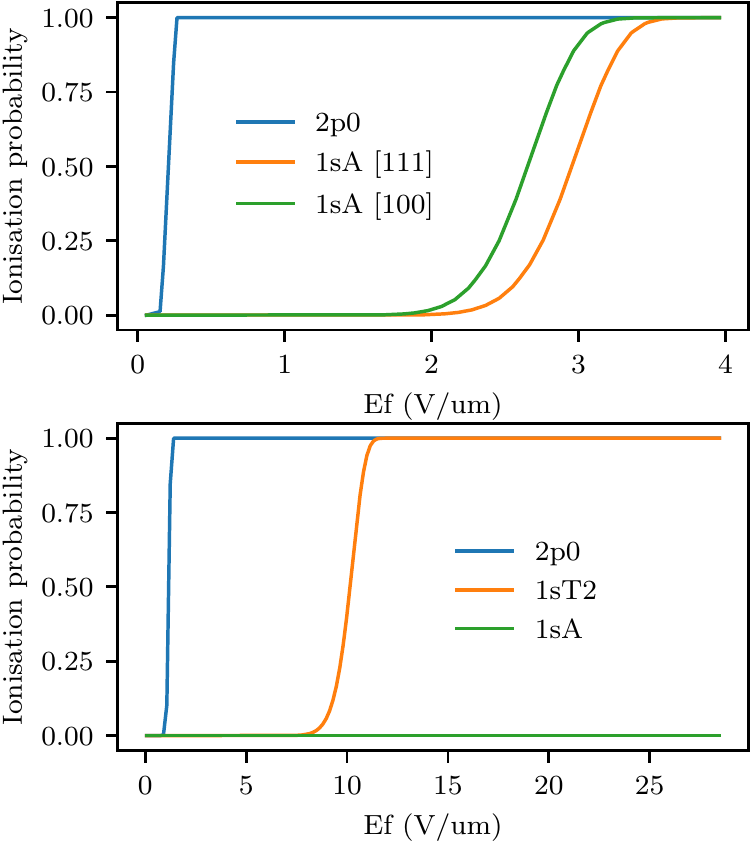}}
\put(-10,240){a)}
\put(-10,120){b)}
\put(70,216){P}
\put(130,93){Se+}
\end{picture}
\caption{\label{fig_ioni}\textbf{Ionisation probability within experimentally determined lifetimes.} a) Ionisation probability within 200ps for various Si:P states and electric field directions. Checked against experimental data from ~\cite{Zurauskiene_1990}. b) Ionisation probability within 7.7ns for theoretical values for Si:Se+ 2p0 (blue),1sT$_2\Gamma_7$ (orange) 1sA (green). Electric field applied along [100].}
\end{figure}

\begin{figure}
\centering
\begin{picture}(220,370)
\put(-10,0){\includegraphics[scale=1]{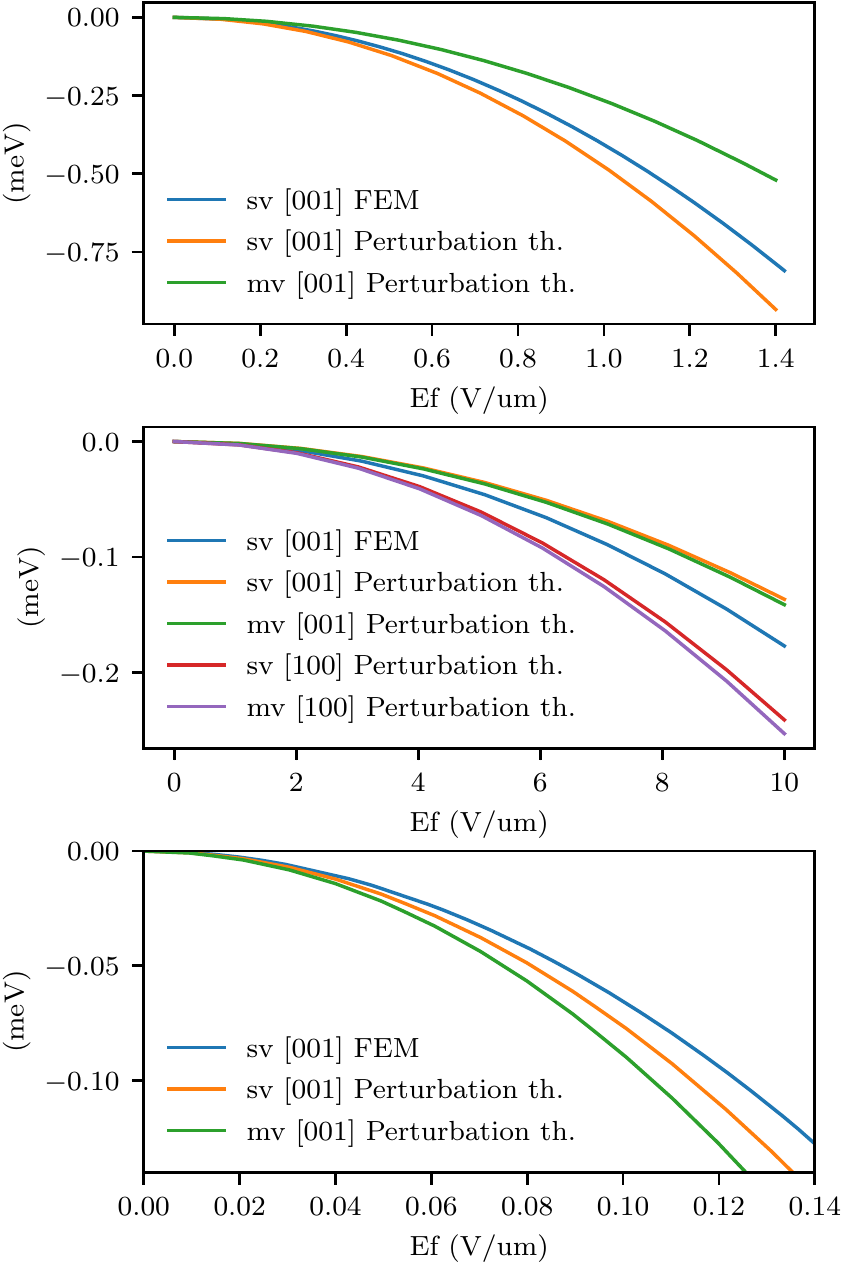}}
\put(-10,355){a)}
\put(-10,235){b)}
\put(-10,115){c)}
\put(45,318){Se+ 2p0}
\put(45,220){Se+ 1sT2}
\put(45,72){P 2p0}
\end{picture}
\caption{\label{fig_stark} \textbf{Stark shifts for $E_f$ smaller than the ionisation limit determined in Fig~\ref{fig_ioni}.} The FEM result includes contributions from the conduction band states not included in the perturbative results. a) Si:Se+ 2p0 Stark shift. c) Si:Se+ 1sT$_2\Gamma_7$ Stark shift. b) Si:P 2p0 Stark shift.}
\end{figure}

\subsection{Ionisation rates}

The ionisation rate for a hydrogen atom in an excited state is given in~\cite{yamabe_theory_1977} to be:
\begin{equation}
    \frac{1}{\tau} = n{-3}[n_2!(n_2 + |m|)!]^{-1}\big(n^3 \frac{E_f}{4}\big)^{-2n_2 -|m|_{-1}}e^{3(n_1 - n_2)\frac{-2}{3n^3 E_f}}\label{eq_yamabe}
\end{equation}
with n, m, $n_1$ and $n_2$ the principal, magnetic, and parabolic quantum numbers respectively (in this paper we consider the 2p0 state which has $n_1=1,n_2=0$) and $E_f$ is the applied field strength in atomic units. We scale the results for the hydrogen atom by using experimental results acquired by \cite{Zurauskiene_1990} on Si:P in the ground state to determine the atomic units for the electric field used in Eq.~\ref{eq_yamabe}. For an atom in the ground state Eq.~\ref{eq_yamabe} reduces to
\begin{equation}
    \frac{1}{\tau} = \frac{\omega \alpha}{F}\exp[-\frac{\alpha}{F}] \label{eq_landau}
\end{equation}
originally given in~\cite{landau1981quantum}, with $\alpha=4 \sqrt{2 m_t} E_f^{3/2}/(3e\hbar)$ the atomic field and $\omega = 12 E_b/\hbar$, where $E_b$ is the binding energy, $e$ is the electron mass and $m_t$ is the tunnelling mass. The tunnelling mass depends on the direction in which is applied the electric field, with the [111] direction yielding the slowest tunnelling times~\cite{Zurauskiene_1990}. The 2D Hamiltonian which enables us to use the FEM for investigating induced dipolar interactions in this paper requires us to consider applying the electric field along [100]. The minimum ionisation rate is given by the Bohr period $\tau_B = 2 \pi \hbar / E_b$~\cite{Zurauskiene_1990}, on the order of 10$^{-13}$ seconds for Si:P and 10$^{-14}$ seconds for Si:Se+.

The results of these formula comply with experimental results for the Si:P 1sA state~\cite{Zurauskiene_1990}. The classical ionisation threshold for Si:P 2p0 is at 0.28 V/um. In order to reduce the ionisation probability during the 2p0 lifetime of 235ps, we can see from Fig.~\ref{fig_ioni} that the electric field should be on the order of 0.2 V/um.

The results for the theoretically calculated ionisation rates for Si:Se+ can be found in Fig.~\ref{fig_ioni}. The 2p0 result was obtained inserting the 1sA $\alpha$ and $\omega$ values into Eq.~\ref{eq_yamabe}. The result for 1sT$_2$ ($n_1$=0, $n_2$=0) coincides with the result obtained when calculating $\alpha$ and $\omega$ values from the binding energy of 1sT$_2$ and directly using Eq.~\ref{eq_landau}. In order to reduce the ionisation probability during the 1sT$_2$ experimentally measured lifetime of 7.7ns, we can see from Fig.~\ref{fig_ioni} that the electric field should be on the order of 8 V/um for 1sT$_2$ and 1V/um for 2p0. In effect, these electric fields yeild negligible dipolar forces, as can be seen from Fig.~\ref{ann_ints}.

\subsection{Stark shifts}

In this paper, we propose a simple way to calculate the exact single valley Stark shift of donors using the FEM, by including a slope corresponding to the electric field in the Schr\"odinger equation in \ref{schr_eq}. This single valley Stark shift enables us to check our multivalley perturbatively calculated Stark shift because it includes contributions from the continuum. This is only possible for a field applied along the polarisation axis as, when applied perpendicular the Schr\"odinger equation no longer reduces to a 2D Hamiltonian.

Multivalley oscillations are on the same scale as the single valley wavefunction for the singly ionised donor Se+, in which the valence electron is tightly pulled in towards the core. All Stark shifts are plotted in Fig.~\ref{fig_stark}, and are very close in magnitude for both methods - single valley or multivalley, for electric fields smaller than the ionising electric fields for the donor excited state lifetimes previously determined.

The perturbatively calculated single valley calculations include single s states with no central cell corrections (ccc) in the 2p0 calculations, and with a ccc for the 1sT$_2\Gamma_7$ calculation in order to be able to compare directly to the FEM results. We can see both results are close, despite the perturbative results not including contributions from the continuum. The multivalley calculations include, for the ground as well as for higher lying s states, A, E and T manifolds, each with their own ccc.

For the 2p0 states of both donors, the main contributing state is 2s in the single valley case, and 2sE and 2sA in the multivalley case. For Se+ 1sT$_2\Gamma_7$ in both single and multivalley cases, the main contributions to the Stark shift parallel to the polarisation of the state come from the p0 states and perpendicular to the polarisation comes from the p+ states.

Considering these Stark shifts, with laser linewidths below 0.1meV and Rabi frequencies on the order of 1meV, small electric fields would suffice to shift the donors on and off resonance. Large fields could even be applied to all the ground state donors not contributing to the gate to fully ensure they are off resonance with the laser exciting to the Rydberg state, because ground state donors can resist far larger fields than the excited states, which have a short lifetime and have higher ionisation probabilities.

\subsection{Stray Electric Fields}
In this section, we show that the stray electric fields from the gates local to the surrounding donors (Stark shifting them off resonance with the laser) do not interfere with the gate operation.

In order to reduce the probability of the laser exciting a transition, the detuning must be increased proportional to the Rabi frequency, for example for a 0.1 probability, $\Delta=4\Omega$. For realistic Rabi frequencies between $10^8$ Hz and $10^{11}$ Hz, depending on the experimental laser power, this sets an upper bound on $\Delta$ of around 0.4 meV.

The 2p0 state in Si:P ionises in relatively small fields as can be seen in the previous section entitled 'Ionisation rates and Stark shifts'. A 0.4meV Stark shift would require an electric field of 0.5 V/$\mu$m, which we calculated using the multivalley perturbation theory method described in that section.

As is shown in the figure in the main text, we use the Finite Element Method to solve for the electric fields in the proposed geometry. We find that if the left local gates of donors not participating in the entangling operation are held at -0.0025V and the right ones at 0.0025V, with the gates surrounding the central donors all held at 0V, and the top global gate at -0.015V and the bottom one at 0.015 V, then the donors not participating in the entangling operation will get shifted off resonance with the laser with a field of 0.5V/$\mu$m, and the two donors participating in the readout, in the case of Si:P 2p0 being chosen as $\ket{r}$, will get a small field parallel to the inter-donor axis due to the global gates, of around 0.18 V$/\mu m$, which will not ionise them within the excited state lifetime, but will induce dipolar interactions. We conclude that with this proposed geometry, there are no unwanted stray electric fields.

\bibliography{apssamp}